\documentclass[aps,prd,a4paper,twocolumn,amsmath,showpacs,superscriptaddress,nofootinbib,preprintnumbers]{revtex4-1}
\pdfoutput=1
\usepackage{amssymb,amsmath,latexsym,mathrsfs}
\usepackage[sort&compress]{natbib}
\usepackage{graphicx,subfigure}
\usepackage{epsfig}
\usepackage{varioref,xr-hyper}
\usepackage{color}
\usepackage{multirow}
\usepackage{array}
\usepackage{hyperref}
\usepackage{wasysym}
\usepackage{color}
\usepackage{float}
\usepackage{xcolor}
\usepackage[utf8]{inputenc}
\usepackage[T1]{fontenc}
\DeclareUnicodeCharacter{2212}{-}
\usepackage[normalem]{ulem}

\newcommand{\Neff}{\ensuremath{N_{\rm eff}}}

\begin{document}


\title{Minimal dark energy: key to sterile neutrino and Hubble constant tensions?}


\author{Eleonora Di Valentino}
\email{e.divalentino@sheffield.ac.uk}
\affiliation{School of Mathematics and Statistics, University of Sheffield, Hounsfield Road, Sheffield S3 7RH, United Kingdom}

\author{Stefano Gariazzo}
\email{gariazzo@to.infn.it}
\affiliation{Istituto Nazionale di Fisica Nucleare (INFN), Sezione di Torino, Via P. Giuria 1, I-10125 Turin, Italy}

\author{Carlo Giunti}
\email{carlo.giunti@to.infn.it}
\affiliation{Istituto Nazionale di Fisica Nucleare (INFN), Sezione di Torino, Via P. Giuria 1, I-10125 Turin, Italy}

\author{Olga Mena}
\email{omena@ific.uv.es }
\affiliation{IFIC, Universidad de Valencia-CSIC, 46071, Valencia, Spain}

\author{Supriya Pan}
\email{supriya.maths@presiuniv.ac.in}
\affiliation{Department of Mathematics, Presidency University, 86/1 College Street,
Kolkata 700073, India}

\author{Weiqiang Yang}
\email{d11102004@163.com}
\affiliation{Department of Physics, Liaoning Normal University, Dalian, 116029, P. R. China}


\begin{abstract}
\emph{Minimal} dark energy models, described by the same number of free parameters of the standard cosmological model with cold dark matter plus a cosmological constant to parameterize the dark energy component, constitute very appealing scenarios which may solve long-standing, pending tensions. On the one hand, they alleviate significantly the tension between cosmological observations and the presence of one sterile neutrino motivated by the short-baseline anomalies: we obtain a $95\%$ {CL }cosmological bound on the mass of a fully thermalized fourth sterile neutrino ($\Neff=4$) equal to $m_s<0.65\ (1.3)$~eV within the PEDE (VM) scenarios under consideration. Interestingly, these limits are in agreement with the observations at short-baseline experiments, and the PEDE scenario is favored with respect to the $\Lambda$CDM case when the full data combination is considered.
On the other hand, the Hubble tension is satisfactorily solved in almost all the \emph{minimal} dark energy
schemes explored here.
These phenomenological scenarios may therefore shed light on differences arising from near and far universe probes, and also on discrepancies between cosmological and laboratory sterile neutrino searches.
\end{abstract}

\maketitle
\section{Introduction}

The standard cosmological model describes surprisingly well the present universe,
which contains mostly dark matter (the ingredient responsible for the growth of the structures we observe today) and dark energy (the largely unknown engine for the current cosmic acceleration).
The most economical nature
of the dark energy component relies on the introduction of a Cosmological Constant (CC) in Einstein’s equations, which has a constant equation of state $w$ with a value $w =-1$.
This assumption leads to the so-called $\Lambda$CDM model~\cite{Planck:2018vyg},
which is the basis of most cosmological analyses.
Nevertheless, when computing the
vacuum energy density of dark energy from a quantum field theory approach,
the expected value for the CC exceeds the measured one by more than 120 orders of magnitude, leading to the so-called \emph{Cosmological Constant Problem} \cite{Weinberg:1988cp}.
In the absence of a fundamental symmetry
which sets the vacuum energy to very small values, it is appropriate to look
for alternative physical mechanisms, such
as quintessence \cite{Wetterich:1987fm,Ratra:1987rm,Ferreira:1997hj,Copeland:1997et,Carroll:1998zi,Zlatev:1998tr,Wetterich:2001jw,Tsujikawa:2013fta}, modified gravity \cite{Capozziello:2002rd,Nojiri:2005vv,Nojiri:2006ri,Sotiriou:2008rp,DeFelice:2010aj,Paliathanasis:2011jq,Capozziello:2011et,Clifton:2011jh,Harko:2011kv,Chakraborty:2012kj,Basilakos:2013rua,Odintsov:2013iba,Cai:2015emx,Paliathanasis:2016vsw,Nunes:2016qyp,Nunes:2016plz,Nunes:2016drj,Nojiri:2017ncd,Nunes:2018evm,Papagiannopoulos:2018mez,Nunes:2019bjq,Paliathanasis:2021ysb,Bahamonde:2021gfp,Paliathanasis:2021uvd}, and other exotic scenarios~\cite{Armendariz-Picon:2000nqq,Armendariz-Picon:2000ulo,Kamenshchik:2001cp,Caldwell:1999ew,Padmanabhan:2002cp,Shapiro:2009dh,Lima:2012cm,Perico:2013mna,Nunes:2016aup,Sola:2016zeg,deHaro:2015hdp,Pan:2016jli,Paliathanasis:2016dhu,Brevik:2017msy,Haro:2015ljc,Pan:2018ibu,Das:2020wfe}.
On the other hand, any extension of the $\Lambda$-cosmology in terms of extra degrees freedom may
increase the reduced $\chi^2$ of the best fit in the extended cosmological scenario compared to the minimal $\Lambda$-cosmology,
and thereby, a question mark is still left in terms of the observational fittings.
Therefore, the construction of a different \emph{minimal} cosmological model, alternative to the standard one with a CC, is very important.
Keeping in mind  the idea of a minimal cosmology, the authors of Ref.~\cite{Li:2019yem} proposed a possible  alternative framework based on phenomenological grounds, namely, the Phenomenologically Emergent Dark Energy (PEDE) model.
On the other hand, the Vacuum Metamorphosis (VM) model, relying on a late-time gravitational phase transition,
is also a possible \emph{minimal} alternative based on the first principles~\cite{Parker:2003as, Parker:2000pr, Caldwell:2005xb}.
These scenarios have exactly the same number of free parameters than the spatially flat $\Lambda$CDM cosmology and they have been shown to provide an excellent solution to the well-known $H_0$ tension; see recent reviews in this direction in Refs.~\cite{DiValentino:2021izs,Perivolaropoulos:2021jda,Schoneberg:2021qvd,Shah:2021onj}.

On another note, in the Standard Model (SM) of elementary particles, from the invisible Z-boson width at the Large Electron Positron collider (LEP) at CERN~\cite{ALEPH:2005ab},
we know that there are three active neutrinos (and antineutrinos) sensitive to weak interactions:
$\nu_e$ ($\bar{\nu}_e$), $\nu_\mu$ ($\bar{\nu}_\mu$) and $\nu_\tau$ ($\bar{\nu}_\tau$).
However, as in the cosmological constant case aforementioned, in which there is no known fundamental symmetry imposing a very small vacuum energy,
there is also no known fundamental reason in nature establishing the number of \emph{sterile} neutrino species, i.e.\ right-handed fermions which do no participate in weak interactions.

Short Baseline (SBL) oscillation experiments~\cite{Gariazzo:2015rra,Giunti:2019aiy,Dasgupta:2021ies}, corresponding to a small ratio of the propagation distance between source and detector and the neutrino energy (e.g.\ distances of $\sim \mathcal{O}(10)$~m for energies of a few MeV), have led to observational anomalies~\cite{LSND:2001aii,MiniBooNE:2007uho,MiniBooNE:2010idf,Bahcall:1994bq,Giunti:2006bj,Giunti:2010zu,Mention:2011rk} that can be explained in a satisfactory way in terms of neutrino oscillations with $\Delta m^2\gtrsim 1$~eV$^2$.
This mass splitting is much larger than the solar and atmospheric ones that emerge from three-neutrino oscillation observations (the atmospheric squared-mass difference being $|\Delta m^2_{31}| \approx 2.55\cdot 10^{-3}$~eV$^2$ and the solar one being $\Delta m^2_{21} \approx 7.5\cdot 10^{-5}$~eV$^2$~\cite{deSalas:2020pgw,Esteban:2020cvm,Capozzi:2021fjo}), which require at least two massive neutrino eigenstates.
Since the sign of the largest mass splitting remains unknown, two mass orderings are possible, called  \emph{normal} ($m_1<m_3$) and \emph{inverted} ($m_3<m_1$).
In the normal ordering, the sum of the neutrino masses must satisfy $\sum m_\nu \gtrsim 0.06$~eV, while in the inverted ordering, $\sum m_\nu \gtrsim 0.10 $~eV, with $\sum m_\nu$
representing the sum of the three light neutrino masses.

In order to explain SBL neutrino oscillation data, one possibility is to invoke the existence of a fourth sterile neutrino state with a mass $m_s \sim 1$~eV~\cite{Abazajian:2012ys,Gariazzo:2017fdh,Dentler:2018sju}, therefore heavier than the active neutrino states\footnote{
Furthermore, a sterile neutrino with even higher mass, at the keV scale, has been invoked to explain other (in this case, astrophysical and cosmological) potential anomalies.
Such a sterile state could significantly contribute to the dark matter content of the universe in the form of a \emph{warm} dark matter-particle,
whose velocity dispersion is neither negligible nor as large as that of neutrinos with energies at the eV scale;
see Refs.~\cite{Dodelson:1993je,Dolgov:2000ew,Abazajian:2001nj,Seljak:2006qw,Boyarsky:2009ix,Viel:2013fqw,Drewes:2016upu,Adhikari:2016bei}.
}.
Short-baseline data hint towards eV sterile neutrino masses and relatively large mixing angles with the active neutrinos~\cite{Melchiorri:2008gq,Kopp:2011qd,Kopp:2013vaa,Gariazzo:2013gua,Gariazzo:2015rra,Gonzalez-Garcia:2015qrr,Gariazzo:2017fdh,IceCube:2016rnb,Giunti:2019aiy,Boser:2019rta,Dentler:2018sju,Hagstotz:2020ukm,steftalk}.
These large mixing angles would ensure the thermalization of sterile neutrinos in the early universe, and since they where relativistic before CMB decoupling, sterile neutrinos contribute to \emph{dark radiation}, parameterized by $\Neff$, which encodes the number of relativistic degrees of freedom in the early universe.
Within the standard models of both particle physics and cosmology, $N_{\rm{eff}}= 3.044$; see the recent Refs.~\cite{Akita:2020szl,Froustey:2020mcq,Bennett:2020zkv} and references therein.
If the light sterile neutrino needed to explain the short baseline anomalies existed in nature, it would be copiously produced in the very early universe via mixing with the active neutrinos,
leading to a value of $\Neff\simeq 4$~\cite{Hannestad:2012ky,Gariazzo:2019gyi}.
Cosmology can therefore provide an independent test to weigh and bound the sterile neutrino masses and abundances, respectively.
Current cosmological bounds on sterile massive species provide
$\Neff < 3.29$ and $m_{\rm eff,\,sterile}<0.65$~eV (assuming $m_s<10$~eV, from Planck TT,TE,EE+lowE+lensing+BAO at $95\%$~CL)~\cite{Planck:2018vyg}~%
\footnote{
Notice that the Planck collaboration reports the constraints on the effective sterile neutrino mass, while in this paper we consider its physical mass instead.
In case of a non-resonant sterile neutrino production, such as in the case of active-sterile neutrino oscillations \cite{Gariazzo:2019gyi},
the two quantities are related by $m_{\rm eff,\,sterile}=m_s\Delta\Neff$,
where $\Delta\Neff$ is the contribution to \Neff\ from the sterile neutrino.
}
(see also Refs.~\cite{Melchiorri:2008gq,Giusarma:2011ex,Dasgupta:2013zpn,Gariazzo:2013gua,Archidiacono:2014apa,Bergstrom:2014fqa,Hagstotz:2020ukm,Seto:2021xua,Seto:2021tad,Feng:2021ipq}),
leaving a little room for the interpretation of the SBL results in terms of a fourth sterile neutrino\footnote{Notice that the limits above assume that the total mass of the active neutrinos is fixed to $\sum m_\nu=0.06$~eV, that is, the minimum mass allowed by oscillation experiments in normal ordering.
In this work, we shall vary the lightest active neutrino mass and compute the limits on that quantity, together with the limits on the sterile neutrino parameters.
The most stringent and recent  $95\%$~CL limits of these parameters are $m_{light}\lesssim 0.025$~eV, $m_s<0.26$~eV and $\Neff < 3.38$~\cite{DiValentino:2021hoh}.
}.
While this tension can be reduced in extensions of the minimal $\Lambda$CDM model including extra interactions and other exotic physics, it is mandatory to test if \emph{minimal} dark energy models can alleviate this controversy.

The question we would like to address here is
whether these \emph{minimal} cosmological scenarios, containing the very same number of degrees of freedom of the canonical $\Lambda$CDM, can alleviate the SBL--cosmology sterile neutrino tension along with a resolution to the Hubble constant tension at the same time. 
We shall explore two possible \emph{minimal} schemes, the Phenomenologically Emergent Dark Energy (PEDE) and the Vacuum Metamorphosis (VM) models, that modify only late time physics, leaving unchanged the standard cosmological history at the inflationary, radiation domination, recombination, CMB decoupling and structure formation periods.
We recall that the aforementioned \emph{minimal} cosmological scenarios have provided an excellent solution to the Hubble constant tension even in presence of the active neutrinos, see Refs.~\cite{Li:2019yem,Pan:2019hac,Yang:2020ope} for PEDE alone and its extension with neutrinos and Refs.~\cite{DiValentino:2017rcr,DiValentino:2020kha,DiValentino:2021zxy} for VM and its various extensions with neutrinos.
The inclusion of the sterile neutrino into these minimal cosmological scenarios, however, is a new ingredient in this article, which significantly differs from their earlier investigations.
To constrain these cosmological scenarios,  we have made use of  a series of observational probes such as Cosmic Microwave Background (CMB) temperature and polarization angular power spectra~\cite{Aghanim:2018eyx,Aghanim:2019ame} and CMB lensing~\cite{Aghanim:2018oex} from Planck 2018, Baryon acoustic oscillations (BAOs) from different astronomical surveys, namely from 6dFGS~\cite{Beutler:2011hx}, SDSS-MGS~\cite{Ross:2014qpa}, and BOSS DR12~\cite{Alam:2016hwk}, 
Pantheon sample~\cite{Scolnic:2017caz} of the Supernovae Type Ia,
a measurement of the Hubble constant from the  SH0ES (Supernovae, $H_0$, for the Equation of State) collaboration~\cite{Riess:2020fzl} and finally SBL neutrino oscillations data, see Section~\ref{sec-data} for details.

The structure of the paper is as follows.
Section~\ref{sec-2} contains a brief description of the two \emph{minimal} dark energy scenarios considered here.
Section~\ref{sec-data} is devoted to explain the cosmological and neutrino oscillation measurements exploited to obtain the results presented and discussed in Section~\ref{sec-results}. We summarize the main findings in Section~\ref{sec-conclusion}.

\section{\emph{Minimal} Dark Energy Models}
\label{sec-2}

According to the observational evidence, our universe is almost homogeneous and isotropic at the largest scales.
This geometric configuration is well described by the  Friedmann-Lema\^{i}tre-Robertson-Walker metric:
\begin{eqnarray}
d{\rm s}^2  = -dt^2 + a^2 (t) \left[\frac{dr^2}{1-kr^2} + r^2 (d\theta^2 + \sin^2 \theta d\phi^2) \right]
\end{eqnarray}
expressed in terms of the comoving coordinates $(t,\; r,\; \theta,\; \phi)$, where $a(t)$ (hereafter we shall use $a$) is the expansion scale factor of the universe. Here, $k$ is a constant representing the curvature of the universe.
For $ k = +1, 0, -1$, we respectively have a closed, flat and open universe.
We assume that the gravitational interaction follows Einstein's general relativity and the matter distribution within the universe is minimally coupled to gravity.
Along with the normal baryons and relativistic radiation, pressureless dark matter, a  dark energy component, we also assume that sterile neutrinos are also present.
Further, we restrict ourselves to the simplest cosmological scenario in which none of the components in the matter sector interacts with any other component.
Under these assumptions, and considering a flat universe (i.e.\ $k =0$), one can write down the Friedmann equations as follows:
\begin{eqnarray}
H^2 = \frac{8 \pi G}{3} \sum_{i} \rho_i\,;\\
2 \dot{H} + 3 H^2 = - 8 \pi G \sum_{i} p_i,
\end{eqnarray}
where $\rho_i$, $p_i$ are respectively the energy density and pressure of the $i$-th fluid component.
Under the assumption of no interactions between any two fluids, all the components enjoy the usual conservation equation, $\dot{\rho}_i + 3 H (1+w_i) \rho_i = 0$, where $w_i = p_i/\rho_i$.
For the dark energy fluid, the conservation equation takes the form $\dot{\rho}_{\rm DE} + 3 H (1+w_{\rm DE}) \rho_{\rm DE} = 0$, where $w_{\rm DE} = p_{\rm DE}/\rho_{\rm DE}$ is the equation-of-state parameter of dark energy.
Solving the conservation equation for dark energy, one gets
\begin{eqnarray}
\rho_{\rm DE} = \rho_{\rm DE,0} \; \exp \left[ 3 \; \int_{0}^{z} \frac{1+w_{DE}(z^\prime)}{1+z^\prime} dz^\prime \right],
\end{eqnarray}
where $\rho_{\rm DE, 0}$ is the present value of the dark energy density.
In the following subsections we describe two \emph{minimal} dark energy models,
which both are descripted by the same number of free parameters as the $\Lambda$CDM framework.

\subsection{Phenomenologically Emergent Dark Energy (PEDE)}

The PEDE model was originally introduced in Ref.~\cite{Li:2019yem} as an alternative to the standard $\Lambda$CDM model.
Although it has exactly the same number of free parameters as the $\Lambda$CDM model, interestingly, in this scenario the DE equation of state has a dynamical character, contrary to the $\Lambda$CDM scenario in which the cosmological constant acting as the DE candidate has a constant equation of state.
Within this model, the Hubble constant value can be increased without any extra degrees of freedom with respect to the cosmological constant case (see also \cite{Pan:2019hac}).
In this new cosmological model, the dark energy density with respect to the current critical energy density is parametrized as
\begin{eqnarray}\label{model}
\frac{\rho_{\rm DE}}{\rho_{\rm crit,0}} = \widetilde{\Omega}_{\rm DE} (z) = \Omega_{\rm DE,0} \Bigl[1- \tanh (\log_{10} (1+z)) \Bigr],
\end{eqnarray}
where $\Omega_{\rm DE,0}$ is the present value of the dark energy density parameter $\Omega_{\rm DE}$.
By construction, this dark energy model has a dynamical evolution, contrary to the time-independent cosmological constant $\Lambda$ in the context of the $\Lambda$CDM model.
Consequently, the cosmological scenario described by this PEDE model has exactly the same number of free parameters than the standard $\Lambda$CDM model, but a different phenomenology.
Making use of the conservation equation for the dark energy component, one could
derive the dark energy equation-of-state as
\begin{eqnarray}\label{de-eos}
w_{\rm DE} (z) =  -1 + \frac{1}{1+z} \times \frac{d\; \ln \widetilde{\Omega}_{\rm DE} (z)}{dz},
\end{eqnarray}
which, for the model in Eq.~(\ref{model}), takes the form:
\begin{align}\label{eos}
w_{\rm DE} (z) = -1 -\frac{1}{3 \ln 10} \times \Bigl[ 1+ \tanh \bigl(\log_{10} (1+z)\bigr) \Bigr]~.
\end{align}
Notice that  $w_{\rm DE}$ asymptotically converges to $-1$ in the far future, while it currently has a phantom nature: $w_{\rm DE} (z=0) =-1 -1/(3\ln 10)$.

\subsection{The Vacuum Metamorphosis model}
\label{ssec:VM}

The Vacuum Metamorphosis (VM) model, inspired on quantum gravitational effects, is based on a late-time gravitational phase transition~\cite{Parker:2003as, Parker:2000pr, Caldwell:2005xb}.
This phase transition will take place when the Ricci scalar curvature $R$ is of the order of the squared mass of the scalar field, $m^2$, which determines the matter density today, $\Omega_m$.
As a consequence, the VM model should be regarded as \emph{minimal}, since it has the same number of free parameters as the flat $\Lambda$CDM scenario.
A further motivation in favor of this model is that it provides a possible solution to the $H_0$ tension~\cite{DiValentino:2017rcr,DiValentino:2020kha,DiValentino:2021zxy}.
The expansion rate before and after the phase transition reads as \cite{DiValentino:2020kha}:
\begin{widetext}
\begin{equation}
\frac{H^2}{H_0^2} =
\begin{cases}
\Omega_m (1+z)^3 \!+\! \Omega_r (1+z)^4  \!+\! \Omega_k (1+z)^2
\!-\! M\left\{1-\left[3\left(\frac{4}{3\Omega_m}\right)^4 M(1-M-\Omega_k-\Omega_r)^3\right]^{-1}
\right\}, &\, \hbox{for $z > z_t^{\rm ph}$}\,;\\
(1-M-\Omega_k)(1+z)^4+\Omega_k(1+z)^2+M\, &\, \hbox{for $z\leq z_t^{\rm ph}$}\,,
\end{cases}
\end{equation}
\end{widetext}
where $M = m^2/(12H_0^2)$ and $\Omega_k=-k/H_0^2$.
The phase transition occurs at the redshift
\begin{equation}
    z_t^{\rm ph} =
    -1+\frac{3\Omega_m}{4(1-M-\Omega_k-\Omega_r)} \,.
\end{equation}
Notice that in the original Vacuum Metamorphosis scenario described above, the mass $m$ of the field, and consequently $M$, are not completely free parameters, but they are related to the matter energy density as follows:
\begin{equation}
\Omega_m=\frac{4}{3}\left[3M(1-M-\Omega_k-\Omega_r)^3\right]^{1/4}~,
\end{equation}
obtained assuming that there is not a cosmological constant at high redshifts. In other words, before the phase transition, the universe behaves as one with matter, radiation, and spatial curvature, while after the phase transition it effectively has a radiation component, that rapidly redshifts away leaving the universe in a
de Sitter phase.
This first principles is therefore not nested with the $\Lambda$CDM scenario.

It is also possible to extend this VM scheme into a more general one, in which $M$ is not further constrained by any relationship and a cosmological constant appears at high redshifts, as the vacuum expectation value of the massive scalar field.
This model is dubbed as \emph{Elaborated Vacuum Metamorphosis} (EMV)~\cite{DiValentino:2017rcr,DiValentino:2020kha} and it is not minimal anymore.
In this case we have to impose the additional conditions $z_t^{\rm ph}\ge0$ and
$\Omega_{\rm DE}(z>z_t^{\rm ph})\ge0$.
In this EVM case the effective equation of state of DE is~\cite{DiValentino:2020kha}:
\begin{widetext}
\begin{equation}
    w_{\rm VMDE}(z)=-1-\frac{1}{3}\frac{3\Omega_m (1+z)^{3}-4(1-M-\Omega_k-\Omega_r)(1+z)^{4}}{M+(1-M-\Omega_k-\Omega_r)(1+z)^{4}-\Omega_m (1+z)^{3}} \,,
    \label{eq:wpde}
\end{equation}
\end{widetext}
after the phase transition, and $w_{\rm VMDE}(z)=-1$ before the phase transition.
Let us remind the reader here that in the case without the cosmological constant,
i.e.\ in the original minimal VM described above, instead,
there is no dark energy before the phase transition.

\section{Observational datasets and methodology}
\label{sec-data}

In this section we describe the cosmological probes and the statistical methodology used to extract the constraints from present cosmological observations,
starting from the cosmological and neutrino oscillation datasets considered in our analyses.

\begin{enumerate}
\item \textbf{Cosmic Microwave Background (CMB)}: The latest
Cosmic microwave background (CMB) temperature and polarization angular power spectra {\it plikTTTEEE+lowl+lowE} from Planck 2018 have been used~\cite{Aghanim:2018eyx,Aghanim:2019ame}.

\item \textbf{Lensing}: The CMB lensing reconstruction likelihood from Planck 2018~\cite{Aghanim:2018oex} has also been used.

\item \textbf{Baryon acoustic oscillation (BAO) distance measurements}: We include Baryon acoustic oscillations (BAOs) from different astronomical surveys, namely from 6dFGS~\cite{Beutler:2011hx}, SDSS-MGS~\cite{Ross:2014qpa}, and BOSS DR12~\cite{Alam:2016hwk}.

\item \textbf{Pantheon sample}:
The Pantheon~\cite{Scolnic:2017caz} sample of the Supernovae Type Ia has been included in the analyses.

\item \textbf{Hubble constant, $H_0$ (R20)}:
A gaussian prior on the Hubble constant from the Supernovae, $H_0$, for the Equation of State (SH0ES) collaboration yielding $H_0 = 73.2 \pm 1.3$ km/sec/Mpc at 68\% CL~\cite{Riess:2020fzl} has also been incorporated.

\item \textbf{SBL neutrino oscillations}:
We consider the posterior probability on $m_s$ obtained with a combined Bayesian
analysis of the data of the $\nu_{\mu}\to\nu_{e}$ appearance experiments
LSND~\cite{LSND:2001aii},
MiniBooNE~\cite{MiniBooNE:2018esg},
BNL-E776~\cite{Borodovsky:1992pn},
KARMEN~\cite{KARMEN:2002zcm},
NOMAD~\cite{NOMAD:2003mqg},
ICARUS~\cite{ICARUS:2013cwr}
and
OPERA~\cite{OPERA:2013wvp}.
We consider only SBL $\nu_{\mu}\to\nu_{e}$ appearance data\footnote{
The current global fit of SBL neutrino oscillation data
does not give reliable results because of a strong tension between
the neutrino oscillation interpretations of appearance and disappearance data
(see the reviews in Refs.~\cite{Giunti:2019aiy,Diaz:2019fwt,Boser:2019rta,Dasgupta:2021ies}).
}
because the positive indication in favor of light sterile neutrinos
given by the LSND and MiniBooNE\footnote{
The recent results of the MicroBooNE experiment~\cite{MicroBooNE:2021zai}
disfavor the interpretation of the MiniBooNE low-energy excess~\cite{MiniBooNE:2018esg} as a photon background~\cite{Hill:2010zy,Giunti:2019sag}
and strengthen the neutrino oscillation interpretation.
} data
is currently the strongest one and has not been directly excluded by other experiments.
The data of the other SBL $\nu_{\mu}\to\nu_{e}$ appearance experiments
constrain the mass and mixings of the sterile neutrino,
leaving an allowed interval of masses around 1 eV
(see Fig.~4a of Ref.~\cite{Giunti:2019aiy}).

\end{enumerate}

Cosmological observables are computed with a modified version of \texttt{CAMB}~\cite{Lewis:1999bs}.
To derive bounds on the proposed scenarios, we modify the efficient and well-known cosmological package \texttt{CosmoMC} \cite{Lewis:2002ah}, publicly available at \url{http://cosmologist.info/cosmomc/}, which we use to constrain the cosmological parameters by means of MCMC sampling.
This cosmological package is equipped with a convergence diagnostic following the Gelman and Rubin prescription~\cite{Gelman:1992zz}.
Additionally, the \texttt{CosmoMC} package supports the Planck 2018 likelihood~\cite{Aghanim:2019ame}.
In Table~\ref{tab:priors} we show the flat priors on the free parameters that we use in the statistical simulations.

Concerning the sterile neutrino sector, we describe it by means of the two usual and well-known parameters, the sterile neutrino mass $m_s$ and the contribution of sterile neutrinos to the effective number of relativistic species, $\Delta N_{\rm{eff}}$. Assuming a sterile neutrino produced in a non-resonant scenario \cite{Dodelson:1993je}, as it is the case when considering neutrino oscillations \cite{Gariazzo:2019gyi}, these two parameters can be combined to define the effective sterile neutrino mass  
$m^{\rm{eff}}_s= \Delta \Neff\, m_s$, which is a measure of the sterile neutrino energy density. Notice that,
in addition to the sterile neutrino parameters $m_s$ and $\Delta N_{\rm{eff}}$, we also consider the mass of the active neutrinos to be a free parameter.
Instead of using $\sum m_\nu$ as a direct parameter in our analyses, we exploit the $m_{light}$ parameter, which, assuming the normal neutrino mass ordering, is related to $\sum m_\nu$ via the solar and atmospheric mass splittings as:

\begin{equation}
\sum m_\nu \equiv m_{light} + \sqrt{m^2_{light}
+\Delta m^2_{21}} + \sqrt{m^2_{light}
+\Delta m^2_{31}}~.
\end{equation}

\begin{table}
\begin{center}
\renewcommand{\arraystretch}{1.4}
\begin{tabular}{|c@{\hspace{1 cm}}|@{\hspace{1 cm}} c|}
\hline
Parameter                    & Prior\\
\hline\hline
$\Omega_{b} h^2$             & $[0.005,0.1]$\\
$\Omega_{c} h^2$             & $[0.001,0.99]$\\
$\tau$                       & $[0.01,0.8]$\\
$n_s$                        & $[0.8, 1.2]$\\
$\log[10^{10}A_{s}]$         & $[1.61,3.91]$\\
$100\theta_{MC}$             & $[0.5,10]$\\
$\Neff$                & $[0.05,10]$\\
$m_{\rm light}$~[eV]         & $[0,5]$\\
$m_s$~[eV]                  & $[0.1,10]$\\
$M_{\rm vacuum}$             & $[0.5,1]$\\
\hline
\end{tabular}
\end{center}
\caption{Flat priors on various free parameters of the different cosmological scenarios explored here.}
\label{tab:priors}
\end{table}

\section{Results and analysis}
\label{sec-results}

\begingroup
\squeezetable
\begin{center}
\begin{table*}
\begin{tabular}{ccccccccccc}
\hline\hline
Parameters & CMB & CMB+BAO & CMB+Lensing & CMB+Pantheon & CMB+R20  \\ \hline
$\Omega_c h^2$ & $    0.1202_{-    0.0019-    0.0037}^{+    0.0018+    0.0040}$ & $    0.1206_{-    0.0016-    0.0038}^{+    0.0018+    0.0037}$ & $    0.1202_{-    0.0017-    0.0035}^{+    0.0017+    0.0038}$ & $    0.1216_{-    0.0018-    0.0040}^{+    0.0018+    0.0038}$ &  $    0.1196_{-    0.0019-    0.0038}^{+    0.0017+    0.0044}$ \\

$\Omega_b h^2$ & $    0.02239_{-    0.00016-    0.00031}^{+    0.00016+    0.00031}$ & $    0.02236_{-    0.00016-    0.00028}^{+    0.00014+    0.00030}$  & $ 0.02240_{-    0.00015-    0.00030}^{+    0.00015+    0.00030}$ & $    0.02227_{-    0.00016-    0.00031}^{+    0.00016+    0.00034}$ & $    0.02246_{-    0.00014-    0.00028}^{+    0.00015+    0.00029}$  \\

$100\theta_{MC}$ & $    1.04078_{-    0.00032-    0.00065}^{+    0.00033+    0.00065}$ & $    1.04073_{-    0.00030-    0.00061}^{+    0.00030+    0.00059}$ & $    1.04078_{-    0.00032-    0.00064}^{+    0.00033+    0.00063}$ & $    1.04051_{-    0.00032-    0.00063}^{+    0.00032+    0.00062}$  & $    1.04090_{-    0.00031-    0.00066}^{+    0.00034+    0.00061}$  \\

 $\tau$ & $    0.0547_{-    0.0080-    0.015}^{+    0.0073+    0.016}$ & $    0.0543_{-    0.0082-    0.015}^{+    0.0073+    0.015}$  & $    0.0547_{-    0.0077-    0.014}^{+    0.0070+    0.015}$  &  $    0.0538_{-    0.0082-    0.014}^{+    0.0070+    0.016}$ &  $    0.0558_{-    0.0080-    0.014}^{+    0.0073+    0.016}$ \\

$n_s$ & $    0.9648_{-    0.0052-    0.0096}^{+    0.0047+    0.0104}$ & $    0.9637_{-    0.0044-    0.0089}^{+    0.0043+    0.0090}$ & $    0.9647_{-    0.0049-    0.0090}^{+    0.0044+    0.0097}$   & $    0.9606_{-    0.0046-    0.0089}^{+    0.0046+    0.0096}$ & $    0.9670_{-    0.0049-    0.0088}^{+    0.0045+    0.0094}$ \\

${\rm{ln}}(10^{10} A_s)$ & $    3.047_{-    0.017-    0.031}^{+    0.015+    0.033}$ & $    3.048_{-    0.017-    0.031}^{+    0.016+    0.033}$  & $    3.048_{-    0.016-    0.029}^{+    0.014+    0.030}$  & $    3.049_{-    0.017-    0.030}^{+    0.015+    0.034}$  & $    3.048_{-    0.016-    0.030}^{+    0.015+    0.032}$  \\

$\Omega_{m0}$ & $    0.287_{-    0.017-    0.027}^{+    0.010+    0.030}$ & $0.2896_{-    0.0083-    0.014}^{+    0.0070+    0.016}$ & $    0.286_{-    0.016-    0.025}^{+    0.010+    0.029}$ & $    0.318_{-    0.021-    0.034}^{+    0.016+    0.036}$  &  $    0.2766_{-    0.0091-    0.017}^{+    0.0076+    0.017}$  \\

$\sigma_8$ & $    0.822_{-    0.016-    0.059}^{+    0.034+    0.047}$ & $    0.821_{-    0.019-    0.048}^{+    0.029+    0.044}$ & $    0.822_{-    0.016-    0.049}^{+    0.030+    0.041}$  & $    0.777_{-    0.038-    0.073}^{+    0.038+    0.073}$  & $    0.833_{-    0.013-    0.045}^{+    0.025+    0.037}$  \\

$H_0$ [km/s/Mpc]& $   71.3_{-    1.0-    2.9}^{+    1.5+    2.5}$ & $   71.04_{-    0.69-    1.50}^{+    0.75+    1.41}$  & $   71.3_{-    0.9-    2.7}^{+    1.4+    2.4}$ & $   68.6_{-    1.4-    3.1}^{+    1.8+    2.9}$  & $   72.25_{-    0.76-    1.5}^{+    0.76+    1.6}$\\

$m_{light}$ [eV]& $  < 0.039 < 0.096 $ & $  < 0.042 < 0.077 $ & $ < 0.040 < 0.086 $  & $ 0.090_{- 0.064}^{+    0.043}, < 0.179 $ & $  < 0.022 < 0.050 $ \\

$\Neff$ & $ < 3.11 < 3.23 $ & $ < 3.12 < 3.24 $ & $ < 3.11 < 3.23 $  & $ < 3.15 < 3.31  $   & $ < 3.11 < 3.25 $  \\


$m_s$ [eV]& $ < 4.68,\, \mbox{unconstr.} $ & $ < 4.73, \,\mbox{unconstr.} $ & $ < 4.69, \, \mbox{unconstr.} $  & $ < 4.26, \,\mbox{unconstr.} $   & $ < 4.97, \,\mbox{unconstr.} $  \\

$S_8$ & $    0.803_{-    0.019-    0.042}^{+    0.022+    0.041}$ & $    0.807_{-    0.018-    0.045}^{+    0.025+    0.042}$ & $    0.803_{-    0.015-    0.034}^{+    0.018+    0.033}$  & $    0.799_{-    0.024-    0.054}^{+    0.028+    0.047}$  & $    0.800_{-    0.016-    0.041}^{+    0.022+    0.038}$  \\

$r_{\rm{drag}}$ [Mpc]& $  146.43_{-    0.35-    1.5}^{+    0.80+    1.2}$ & $  146.30_{-    0.32-    1.5}^{+    0.76+    1.1}$ & $  146.43_{-    0.34-    1.5}^{+    0.81+    1.2}$ & $  145.9_{-    0.4-    2.0}^{+    1.1+    1.4}$ & $ 146.59_{-    0.30-    1.6}^{+    0.80+    1.2}$ \\

$r_{\rm drag} h$ [Mpc] & $  104.4_{-    1.6-    4.4}^{+    2.4+    4.0}$ & $  103.9_{-    1.1-    2.3}^{+    1.2+    2.1}$  & $  104.5_{-    1.6-    4.2}^{+    2.4+    3.8}$  & $  100.1_{-    2.3-    4.6}^{+    2.5+    4.3}$  & $  105.9_{-    1.3-    2.7}^{+    1.4+    2.5}$  \\

\hline\hline
\end{tabular}
\caption{Constraints on various free and derived parameters within the PEDE scenario. We report 68\% and 95\% CL limits. When the upper limit coincides with the upper prior range (see Tab.~\ref{tab:priors}), as in the case of $m_s$, we mark the parameter as ``unconstrained''. }\label{tab:PEDE1}
\end{table*}
\end{center}
\endgroup
\begingroup
\squeezetable
\begin{center}
\begin{table*}
\begin{tabular}{ccccccccc}
\hline\hline

Parameters & CMB+Lensing+BAO&~~~ CMB+Lensing+BAO & CMB+Lensing+BAO&~~~ CMB+Lensing+BAO \\
& +Pantheon ($\Neff$ free) & +Pantheon+SBL ($\Neff$ free) &~~~ +Pantheon ($\Neff = 4$) &~~~ +Pantheon+SBL ($\Neff = 4$)  \\

\hline

$\Omega_c h^2$ & $    0.1210_{-    0.0015-    0.0037}^{+    0.0017+    0.0031}$& $    0.1226_{-    0.0018-    0.0028}^{+    0.0012+    0.0031}$ & $    0.1353_{-    0.0012-    0.0025}^{+    0.0013+    0.0024}$ & $    0.1337_{-    0.0011-    0.0022}^{+    0.0011+    0.0021}$\\

$\Omega_b h^2$ & $    0.02233_{-    0.00014-    0.00027}^{+    0.00014+    0.00029}$ & $    0.02235_{-    0.00015-    0.00028}^{+    0.00014+    0.00030}$ & $    0.02293_{-    0.00013-    0.00026}^{+    0.00014+    0.00026}$ & $    0.02298_{-    0.00014-    0.00026}^{+    0.00014+    0.00027}$ \\

$100\theta_{MC}$ & $    1.04066_{-    0.00029-    0.00059}^{+    0.00029+    0.00058}$ & $    1.04056_{-    0.00030-    0.00064}^{+    0.00031+    0.00058}$ & $    1.03934_{-    0.00029-    0.00056}^{+    0.00029+    0.00056}$ & $    1.03949_{-    0.00029-    0.00055}^{+    0.00028+    0.00056}$ \\

$\tau$ & $    0.0538_{-    0.0079-    0.015}^{+    0.0072+    0.016}$ & $    0.0543_{-    0.0079-    0.015}^{+    0.0072+    0.015}$ & $0.0628_{-    0.0094-    0.016}^{+    0.0077+    0.018}$ & $0.068_{-    0.010-    0.016}^{+    0.008+    0.018}$ \\

$n_s$ & $    0.9623_{-    0.0044-    0.0079}^{+    0.0040+    0.0086}$  & $    0.9652_{-    0.0045-    0.0085}^{+    0.0044+    0.0092}$ & $    0.9926_{-    0.0037-    0.0075}^{+    0.0038+    0.0073}$ & $    0.9922_{-    0.0036-    0.0070}^{+    0.0036+    0.0072}$ \\

${\rm{ln}}(10^{10} A_s)$ & $    3.048_{-    0.016-    0.030}^{+    0.015+    0.031}$ & $    3.050_{-    0.017-    0.030}^{+    0.015+    0.031}$ & $    3.094_{-    0.018-    0.033}^{+    0.015+    0.034}$ & $    3.102_{-    0.019-    0.031}^{+    0.016+    0.036}$  \\

$\Omega_{m0}$ &   $   0.2976_{-    0.0075-    0.015}^{+    0.0075+    0.015}$ &   $   0.2972_{-    0.0084-    0.016}^{+    0.0074+    0.015}$ & $  0.2965_{-    0.0083-    0.016}^{+    0.0080+    0.016}$ & $  0.3062_{-    0.0064-    0.013}^{+    0.0064+    0.013}$ \\

$\sigma_8$ & $    0.809_{-    0.018-    0.042}^{+    0.023+    0.040}$ & $    0.821_{-    0.016-    0.032}^{+    0.016+    0.032}$ &  $    0.815_{-    0.021-    0.045}^{+    0.025+    0.042}$ &  $    0.775_{-    0.011-    0.023}^{+    0.012+    0.022}$ \\

$H_0$ [km/s/Mpc]& $   70.32_{-    0.69-    1.5}^{+    0.78+    1.4}$ & $   70.52_{-    0.80-    1.6}^{+    0.89+    1.5}$ & $   74.29_{-    0.85-    1.5}^{+    0.79+    1.6}$ & $   73.30_{-    0.56-    1.1}^{+    0.57+    1.1}$  \\

$m_{light}$ [eV]& $ 0.047_{-    0.031}^{+    0.025}, < 0.092 $ & $ 0.051_{-    0.028}^{+    0.027}, < 0.096 $ & $   < 0.045 < 0.123 $ & $   < 0.015 < 0.033 $ \\

$\Neff$ & $ < 3.13 < 3.23 $ & $ < 3.17 < 3.33 $ & [$4$] & [$4$] \\

$m_s$ [eV]& $ < 4.44, \,\mbox{unconstr.}    $ & $ 0.75^{+0.06+0.17}_{-0.10-0.17}    $ & $ < 0.47 < 0.65 $ & $ 0.678^{+0.066+0.11}_{-0.052-0.12}    $ \\

$S_8$ & $    0.806_{-    0.015-    0.038}^{+    0.021+    0.035}$ & $    0.817_{-    0.013-    0.026}^{+    0.013+    0.025}$ & $    0.809_{-    0.017-    0.034}^{+    0.017+    0.032}$ & $    0.783_{-    0.012-    0.023}^{+    0.012+    0.023}$ \\

$r_{\rm{drag}}$ [Mpc]& $  146.16_{-    0.36-    1.4}^{+    0.79+    1.1}$ & $  145.83_{-    0.5-    1.9}^{+    1.1+    1.4}$ & $  138.19_{-    0.21-    0.41}^{+    0.21+    0.43}$ & $  138.18_{-    0.21-    0.41}^{+    0.21+    0.42}$ \\

$r_{\rm drag} h$ [Mpc]& $  102.8_{-    1.1-    2.1}^{+    1.1+    2.1}$ & $  102.8_{-    1.1-    2.1}^{+    1.1+    2.1}$ & $  102.7_{-    1.2-    2.2}^{+    1.2+    2.3}$ & $  101.29_{-    0.87-    1.7}^{+    0.87+    1.7}$  \\
\hline\hline
\end{tabular}
\caption{Constraints on various free and derived parameters within the PEDE scenario, considering the combined datasets CMB+Lensing+BAO+Pantheon and CMB+Lensing+BAO+Pantheon+SBL. We present the cases where $\Neff$ is either freely varying (second and third columns), or fixed to $\Neff = 4$ (fourth and fifth columns). }\label{tab:PEDE2}
\end{table*}
\end{center}
\endgroup

We start here discussing the results within the PEDE scenario, and comparing them to those in other possible cosmologies.
In Tables \ref{tab:PEDE1} and \ref{tab:PEDE2} we have summarized the constraints obtained from all possible datasets and their combinations.
From the results depicted in Tab.~\ref{tab:PEDE1}, we note that the limits on the sterile neutrino mass and also on the lightest active neutrino mass are mildly relaxed compared to the $\Lambda$CDM constraints shown in Tab.~\ref{tab:LCDM0}. 
This is a consequence of the negative value adopted by the dark energy equation of state within the PEDE framework, which is $w_{\rm DE}<-1$ at very low redshifts (see Eq.~(\ref{de-eos})),
and of the well-known strong degeneracy between the dark energy equation of state and the neutrino mass~\cite{Hannestad:2005gj}.
If $w_{\rm DE}$ is allowed to vary, $\Omega_{m0}$
can take very high values, as required when either the active or the sterile neutrino masses are increased.
For all the data combinations shown in Tab.~\ref{tab:PEDE1}, the sterile neutrino mass is unconstrained at $95\%$~CL:
therefore, the strong existing tension between cosmological bounds and the SBL fits is robustly alleviated.
From the results shown in Tab.~\ref{tab:PEDE1}, one notes that not only the cosmological sterile neutrino mass bounds are weakened in the context of \emph{minimal} dark energy models such as the PEDE one, but also the bounds  on the active neutrino masses are considerably relaxed with respect to the ones obtained in the $\Lambda$CDM scheme.
The combination of CMB and BAO observations provides a $95\%$~CL limit of $m_{light}<0.077$~eV, while within the standard cosmological picture the corresponding bound is $m_{light}<0.041$~eV for the very same data combination.
Interestingly, in one case, there is a (very) mild indication of a non-zero value for the lightest neutrino mass ($m_{light}= 0.090_{-0.064}^{+0.043}$~eV at $68\%$~CL when combining CMB plus SNIa Pantheon data).
These findings agree with previous results in the literature; see e.g.~\cite{Yang:2020ope}, where an indication for a non-zero neutrino mass with a significance of $2\sigma$ ($\sum m_\nu= 0.21^{+0.15}
_{−0.14}$~eV) was found, always in the context of the PEDE scenario\footnote{This bound results from the combination of CMB+BAO+Pantheon+Lensing data plus a prior on the Hubble constant plus Dark Energy Survey (DES) measurements.}.

Within the PEDE model, the values of $N_{\rm{eff}}$ are not shifted to larger values, as the $95\%$~CL upper bounds on the effective number of relativistic degrees of freedom are very close to those computed within the canonical $\Lambda$CDM picture.
For instance, $N_{\rm{eff}}<3.24$~eV at $95\%$~CL for the CMB plus BAO data analyses.
Therefore, one should still advocate some mechanism to suppress the thermalization of the sterile neutrino state in the early universe and its contribution to $N_{\rm{eff}}$; see e.g.~\cite{Bergstrom:2014fqa}.
Among the large number of possible scenarios proposed in the literature, an incomplete list of possibilities includes
exotic sterile neutrino interactions~\cite{Bento:2001xi,Dasgupta:2013zpn,Hannestad:2013ana,Archidiacono:2014nda,Chu:2015ipa,Archidiacono:2016kkh,Song:2018zyl,Archidiacono:2020yey},
a dark energy model with a redshift dependent equation of state
\cite{Giusarma:2011zq},
low reheating temperature scenarios~\cite{Gelmini:2004ah}
and lepton asymmetries~\cite{Foot:1995bm,Chu:2006ua,Saviano:2013ktj}.

In addition to the relief provided on the sterile neutrino tension, the PEDE scenario also offers a dark energy solution that alleviates the $H_0$ tension~\cite{DiValentino:2021izs}.
Notice that within the PEDE \emph{minimal} cosmology  the value of $H_0$ for Planck 2018 CMB + Lensing ($H_0= 71.33_{-0.93}^{+1.44}$~km s$^{−1}$ Mpc$^{−1}$ at 68$\%$~CL) is much larger than the one obtained within the $\Lambda$CDM framework ($H_0= 66.5_{- 0.7}^{+    1.0}$~km s$^{−1}$ Mpc$^{−1}$ at also 68$\%$~CL); see previous findings in  Refs.~\cite{Li:2019yem,Pan:2019hac,Rezaei:2020mrj,Yang:2020ope}.
It is crucial to stress that the larger value of the Hubble constant obtained in PEDE models is robust against different data combinations.

\begin{figure}[t]
    \begin{center}
    \includegraphics[width=0.5\textwidth]{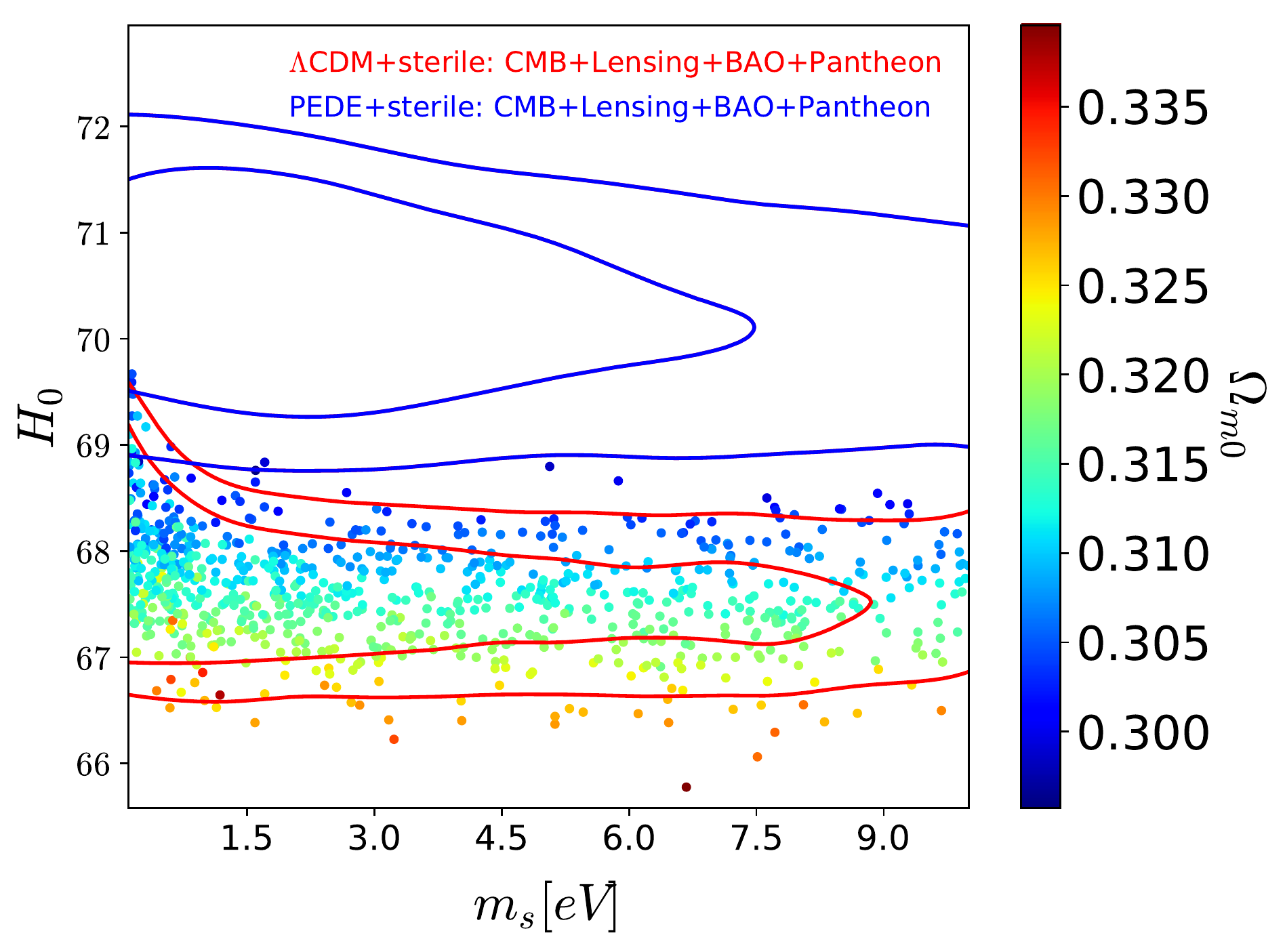}
    \includegraphics[width=0.5\textwidth]{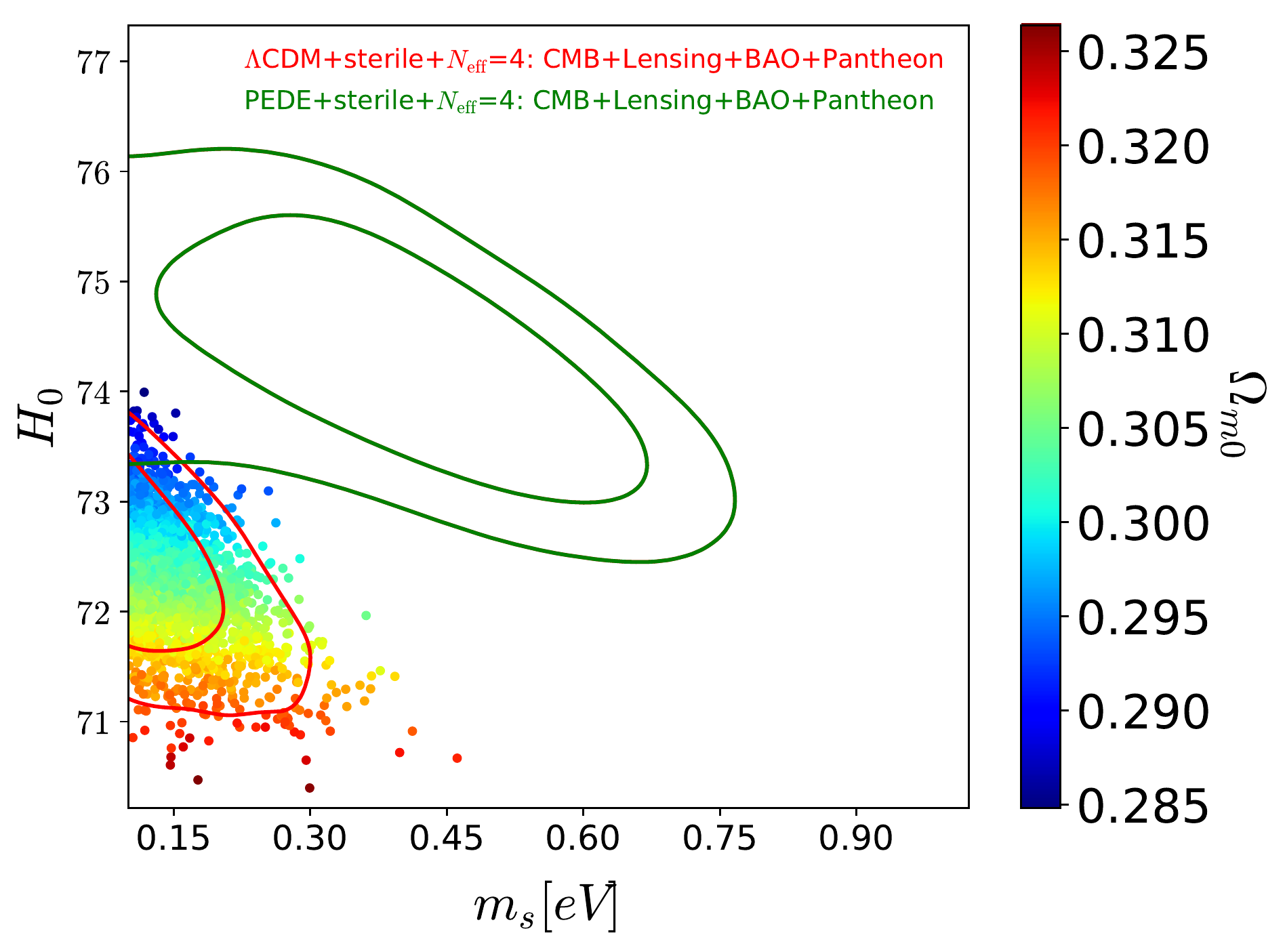}
     \end{center}
    \caption{Upper panel: Samples from CMB+Lensing+BAO+Pantheon chains in the $m_s-H_0$ plane, colour-coded by $\Omega_{m0}$ within the $\Lambda$CDM picture.
    The red (blue) solid contours  show the $68\%$ and $95\%$~CL constraints from  CMB+Lensing+BAO+Pantheon data in the $\Lambda$CDM (PEDE) scenario.
    The parameter $\Neff$ has been assumed to be a free parameter. Lower Panel: Samples from CMB + Lensing + BAO + Pantheon chains in the $m_s-H_0$ plane, colour-coded by $\Omega_{m0}$ within the $\Lambda$CDM picture.
    The red (green) solid contours show the $68\%$ and $95\%$~CL constraints from CMB+Lensing+BAO+Pantheon data in the $\Lambda$CDM (PEDE) scenario.
    In this case, $\Neff=4$. }
    \label{fig:3D}

\end{figure}
\begin{figure}[t]
    \begin{center}
    \includegraphics[width=0.5\textwidth]{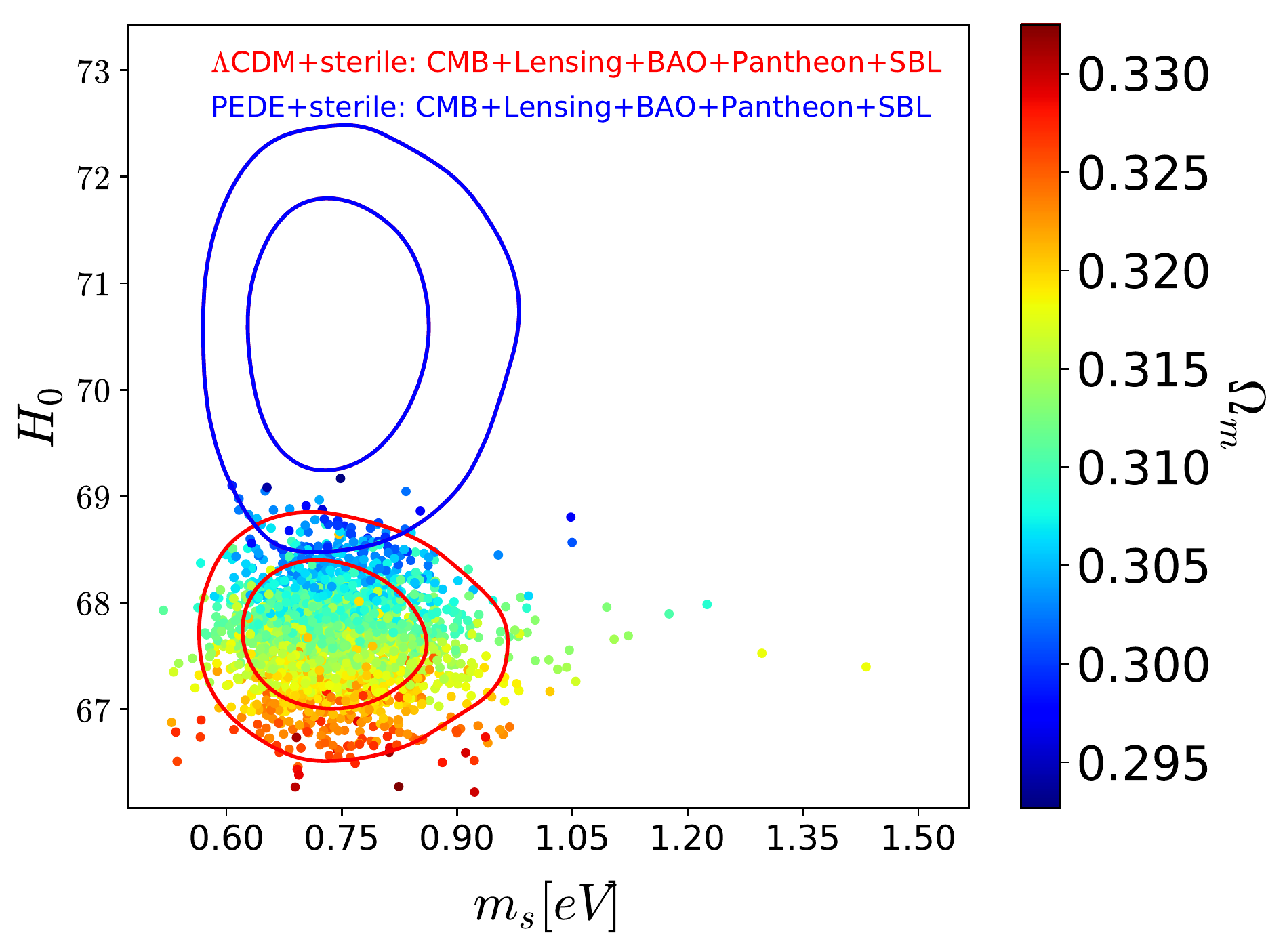}
    \includegraphics[width=0.5\textwidth]{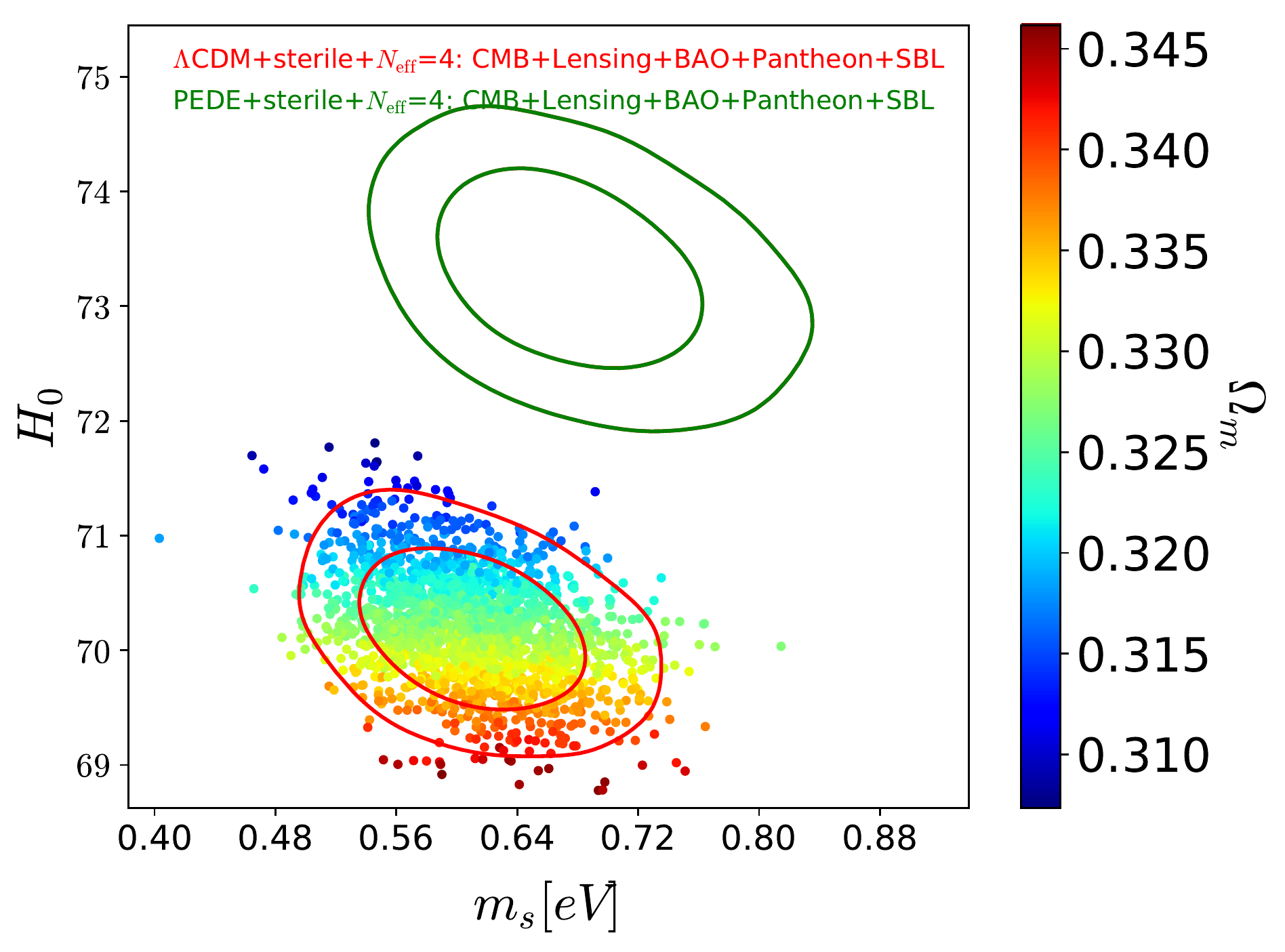}
     \end{center}
    \caption{Upper panel: Samples from CMB+Lensing+BAO+Pantheon+SBL chains in the $m_s-H_0$ plane, colour-coded by $\Omega_{m0}$ within the $\Lambda$CDM picture.
    The red (blue) solid contours  show the $68\%$ and $95\%$~CL constraints from  CMB+Lensing+BAO+Pantheon+SBL data in the $\Lambda$CDM (PEDE) scenario.
    The parameter $\Neff$ has been assumed to be a free parameter. Lower Panel: Samples from CMB+Lensing+BAO+Pantheon+SBL chains in the $m_s-H_0$ plane, colour-coded by $\Omega_{m0}$ within the $\Lambda$CDM picture.
    The red (green) solid contours show the $68\%$ and $95\%$~CL constraints from CMB+Lensing+BAO+Pantheon+SBL data in the $\Lambda$CDM (PEDE) scenario.
    In this case, $\Neff=4$.}
    \label{fig:3D-2}

\end{figure}

\begingroup                                                                                                                     
\squeezetable                                                                                                                   
\begin{center}                                                                                                                  
\begin{table*}                                              
\begin{tabular}{cccccccccccc}       
\hline\hline                                                 
Parameters & CMB & CMB+BAO & CMB+Lensing & CMB+Pantheon & CMB+R20 \\ \hline

$\Omega_c h^2$ & $    0.1202_{-    0.0017-    0.0038}^{+    0.0018+    0.0036}$ & $    0.1190_{-    0.0013-    0.0031}^{+    0.0014+    0.0032}$  & $    0.1204_{-    0.0017-    0.0039}^{+    0.0017+    0.0038}$  & $    0.1197_{-    0.0016-    0.0035}^{+    0.0016+    0.0035}$  & $    0.1219_{-    0.0038-    0.0059}^{+    0.0030+    0.0062}$ \\

$\Omega_b h^2$ & $    0.02237_{-    0.00016-    0.00030}^{+    0.00015+    0.00030}$ & $    0.02246_{-    0.00015-    0.00028}^{+    0.00014+    0.00028}$  & $    0.02237_{-    0.00015-    0.00030}^{+    0.00015+    0.00029}$  & $    0.02240_{-    0.00015-    0.00030}^{+    0.00015+    0.00031}$ & $    0.02274_{-    0.00017-    0.00034}^{+    0.00017+    0.00034}$ \\

$100\theta_{MC}$ & $    1.04077_{-    0.00033-    0.00065}^{+    0.00034+    0.00066}$ & $    1.04097_{-    0.00030-    0.00060}^{+    0.00030+    0.00058}$   & $    1.04076_{-    0.00032-    0.00064}^{+    0.00032+    0.00065}$  & $    1.04086_{-    0.00030-    0.00066}^{+    0.00034+    0.00061}$  & $    1.04073_{-    0.00043-    0.00085}^{+    0.00043+    0.00084}$ \\

$\tau$ & $    0.0548_{-    0.0086-    0.016}^{+    0.0074+    0.016}$ & $    0.0558_{-    0.0082-    0.015}^{+    0.0074+    0.016}$ & $    0.0557_{-    0.0082-    0.014}^{+    0.0071+    0.016}$  & $    0.0556_{-    0.0086-    0.015}^{+    0.0072+    0.017}$ & $    0.0599_{-    0.0091-    0.016}^{+    0.0082+    0.018}$ \\

$n_s$ & $    0.9642_{-    0.0047-    0.0092}^{+    0.0048+    0.0095}$ & $    0.9678_{-    0.0041-    0.0084}^{+    0.0040+    0.0083}$ & $    0.9641_{-    0.0046-    0.0090}^{+    0.0046+    0.0095}$ & $    0.9658_{-    0.0048-    0.0087}^{+    0.0045+    0.0090}$ & $    0.9793_{-    0.0076-    0.0122}^{+    0.0066+    0.0131}$ \\

${\rm{ln}}(10^{10} A_s)$ & $    3.048_{-    0.018-    0.033}^{+    0.015+    0.034}$ & $    3.046_{-    0.017-    0.031}^{+    0.015+    0.033}$  & $    3.050_{-    0.016-    0.029}^{+    0.015+    0.032}$ & $    3.048_{-    0.018-    0.031}^{+    0.015+    0.035}$  & $    3.061_{-    0.019-    0.037}^{+    0.018+    0.039}$ \\

$\Omega_{m0}$ & $    0.330_{-    0.016-    0.027}^{+    0.010+    0.031}$ &  $    0.3142_{-    0.0070-    0.012}^{+    0.0064+    0.013}$ & $    0.330_{-    0.016-    0.025}^{+    0.010+    0.028}$  & $    0.322_{-    0.012-    0.020}^{+    0.0088+    0.021}$  & $    0.3033_{-    0.0089-    0.015}^{+    0.0078+    0.017}$ \\

$\sigma_8$ & $    0.780_{-    0.015-    0.052}^{+    0.030+    0.042}$ & $    0.792_{-    0.010-    0.032}^{+    0.018+    0.028}$ & $    0.783_{-    0.014-    0.044}^{+    0.026+    0.036}$   & $    0.788_{-    0.013-    0.040}^{+    0.023+    0.033}$  & $    0.802_{-    0.011-    0.025}^{+    0.013+    0.024}$ \\

$H_0$ & $   66.4_{-    0.7-    2.2}^{+    1.1+    2.0}$ & $   67.6_{-    0.5-    1.0}^{+    0.5+    1.1}$ & $   66.5_{-    0.7-    2.0}^{+    1.0+    1.8}$ & $   67.0_{-    0.7-    1.6}^{+    0.8+    1.4}$ & $   69.4_{-    1.1-    1.9}^{+    0.9+    1.9}$ \\

$m_{light}$ & $  < 0.038 < 0.094  $ & $ < 0.018 < 0.041 $ & $ < 0.035 < 0.079  $  & $ < 0.027 < 0.060 $  & $ < 0.014 < 0.032 $ \\

$N_{\rm eff}$ & $ < 3.11 < 3.21  $ &  $  < 3.09 < 3.17 $ & $ < 3.11 < 3.23 $  & $  < 3.10 < 3.18  $  & $ < 3.40 < 3.61 $ \\

$m_s$ [eV] & $ < 4.82, uncons.   $ & $  uncons. $ & $ < 5.01, uncons. $ & $ < 5.07, uncons.  $  & $ < 0.38 < 1.11   $ \\

$S_8$ & $    0.818_{-    0.018-    0.043}^{+    0.023+    0.041}$ & $    0.811_{-    0.014-    0.035}^{+    0.018+    0.034}$ & $    0.820_{-    0.015-    0.037}^{+    0.019+    0.034}$ & $    0.816_{-    0.018-    0.041}^{+    0.020+    0.037}$ & $    0.807_{-    0.016-    0.031}^{+    0.016+    0.033}$  \\


$r_{\rm{drag}}$ & $  146.47_{-    0.36-    1.4}^{+    0.72+    1.1}$ & $  146.89_{-    0.26-    1.0}^{+    0.52+    0.9}$ & $  146.41_{-    0.35-    1.5}^{+    0.79+    1.2}$ & $  146.66_{-    0.32-    1.2}^{+    0.62+    1.0}$ & $  144.9_{-    1.5-    3.2}^{+    2.2+    3.0}$ \\


$r_{\rm drag} h$ [Mpc] & $   97.3_{-    1.3-    3.6}^{+    1.9+    3.2}$ & $   99.2_{-    0.8-    1.6}^{+    0.8+    1.6}$ & $   97.4_{-    1.3-    3.2}^{+    1.8+    2.9}$  & $   98.2_{-    1.1-    2.5}^{+    1.4+    2.4}$ & $  100.6_{-    1.1-    2.2}^{+    1.1+    2.1}$ \\

\hline\hline                                            
\end{tabular}                                                                                            
\caption{Constraints on various free and derived parameters within the $\Lambda$CDM scenario. We report 68\% and 95\% CL limits. When the upper limit coincides with the upper prior range (see Tab.~\ref{tab:priors}), as in the case of $m_s$, we mark the parameter 
as ``unconstrained''. }
\label{tab:LCDM0}                                                                                                   
\end{table*}                                                                                                                     
\end{center}                                                                                                                    
\endgroup    

\begingroup
\squeezetable
\begin{center}
\begin{table*}
\begin{tabular}{cccccccccc}
\hline\hline
Parameters & CMB+Lensing+BAO&~~~ CMB+Lensing+BAO & CMB+Lensing+BAO&~~~ CMB+Lensing+BAO \\
& +Pantheon ($\Neff$ free) & +Pantheon+SBL ($\Neff$ free) &~~~ +Pantheon ($\Neff = 4$) &~~~ +Pantheon+SBL ($\Neff = 4$)  \\
\hline

$\Omega_c h^2$ & $    0.1190_{-    0.0015-    0.0031}^{+    0.0013+    0.0033}$ & $    0.11976_{-    0.0012-    0.0022}^{+    0.00098+    0.0022}$ & $    0.1335_{-    0.0011-    0.0021}^{+    0.0010+    0.0021}$ & $    0.1311_{-    0.0011-    0.0022}^{+    0.0011+    0.0022}$  \\

$\Omega_b h^2$ & $    0.02247_{-    0.00014-    0.00028}^{+    0.00014+    0.00028}$ & $    0.02248_{-    0.00014-    0.00028}^{+    0.00014+    0.00028}$  & $    0.02311_{-    0.00014-    0.00027}^{+    0.00014+    0.00026}$ & $    0.02319_{-    0.00014-    0.00027}^{+    0.00013+    0.00028}$ \\

$100\theta_{MC}$ & $    1.04099_{-    0.00030-    0.00059}^{+    0.00030+    0.00056}$ & $    1.04093_{-    0.00031-    0.00059}^{+    0.00031+    0.00060}$  & $    1.03956_{-    0.00029-    0.00057}^{+    0.00029+    0.00055}$ & $    1.03977_{-    0.00029-    0.00057}^{+    0.00029+    0.00056}$\\

$\tau$ & $    0.0585_{-    0.0083-    0.015}^{+    0.0071+    0.015}$ & $    0.0584_{-    0.0083-    0.015}^{+    0.0071+    0.016}$ & $    0.0656_{-    0.0097-    0.017}^{+    0.0079+    0.018}$ & $    0.080_{-    0.010-    0.019}^{+    0.010+    0.020}$ \\

$n_s$ & $    0.9678_{-    0.0044-    0.0078}^{+    0.0038+    0.0086}$ & $    0.9688_{-    0.0041-    0.0078}^{+    0.0041+    0.0083}$ & $    0.9985_{-    0.0037-    0.0075}^{+    0.0038+    0.0076}$ & $    0.9992_{-    0.0038-    0.0076}^{+    0.0039+    0.0077}$ \\

${\rm{ln}}(10^{10} A_s)$ & $    3.052_{-    0.017-    0.028}^{+    0.014+    0.031}$ & $    3.053_{-    0.016-    0.028}^{+    0.014+    0.032}$  & $    3.097_{-    0.018-    0.033}^{+    0.015+    0.034}$ & $    3.121_{-    0.019-    0.036}^{+    0.019+    0.038}$ \\

$\Omega_{m0}$ & $    0.3135_{-    0.0067-    0.012}^{+    0.0059+    0.012}$ & $    0.3134_{-    0.0064-    0.011}^{+    0.0059+    0.012}$  & $    0.3051_{-    0.0068-    0.012}^{+    0.0062+    0.013}$ & $    0.3281_{-    0.0064-    0.012}^{+    0.0064+    0.013}$ \\

$\sigma_8$ & $    0.7965_{-    0.0090-    0.027}^{+    0.015+    0.024}$ & $    0.8007_{-    0.0077-    0.018}^{+    0.010+    0.017}$   & $    0.8194_{-    0.0094-    0.025}^{+    0.014+    0.022}$ & $0.746_{-    0.010-    0.0200}^{+    0.010+    0.019}$ \\

$H_0$ & $   67.6_{-    0.48-    0.96}^{+    0.47+    1.04}$ & $   67.7_{-    0.45-    0.92}^{+    0.46+    0.89}$ & $   72.2_{-    0.53-    1.1}^{+    0.58+    1.05}$ & $   70.1_{-    0.48-    0.94}^{+    0.48+    0.96}$ \\

$m_{light}$ [eV] & $ < 0.016 < 0.036  $ & $ < 0.016 < 0.034  $  & $ < 0.021 < 0.044  $  & $ < 0.010 < 0.022  $  \\

$N_{\rm eff}$ & $ < 3.088 < 3.12 $ & $ < 3.11 < 3.20 $ & [4] & [4] \\

$m_s$ [eV] & $ < 5.11, uncons. $ & $    0.775_{-    0.107-    0.183}^{+    0.068+    0.190}$  & $<0.167, <0.255$ & $    0.621_{-    0.056-    0.102}^{+    0.048+    0.104}$ \\

$S_8$ & $    0.814_{-    0.012-    0.028}^{+    0.015+    0.027}$ & $    0.818_{-    0.011-    0.022}^{+    0.011+    0.021}$ & $    0.826_{-    0.012-    0.023}^{+    0.012+    0.023}$ & $    0.780_{-    0.012-    0.023}^{+    0.011+    0.022}$ \\

$r_{\rm{drag}}$ & $  146.9_{-    0.2-    1.2}^{+    0.6+    0.9}$ & $  146.7_{-    0.3-    1.1}^{+    0.6+    0.9}$  & $  138.6_{-    0.2-    0.4}^{+    0.2+    0.4}$ & $  138.6_{-    0.2-    0.4}^{+    0.2+    0.43}$ \\


$r_{\rm drag} h$ [Mpc] & $   99.3_{-    0.8-    1.5}^{+    0.8+    1.5}$ & $   99.3_{-    0.8-    1.5}^{+    0.8+    1.5}$  & $  100.1_{-    0.8-    1.7}^{+    0.8+    1.6}$ & $   97.2_{-    0.8-    1.5}^{+    0.8+    1.5}$ \\

\hline\hline
\end{tabular}
\caption{Constraints on various free and derived parameters within the $\Lambda$CDM scenario, the combined datasets CMB+Lensing+BAO+Pantheon and CMB+Lensing+BAO+Pantheon+SBL. We present the cases where $\Neff$ is either freely varying (second and third columns), or fixed to $\Neff = 4$ (fourth and fifth columns).
}
\label{tab:LCDM}
\end{table*}
\end{center}
\endgroup

\begin{figure}[t]
    \centering
    \includegraphics[width=0.5\textwidth]{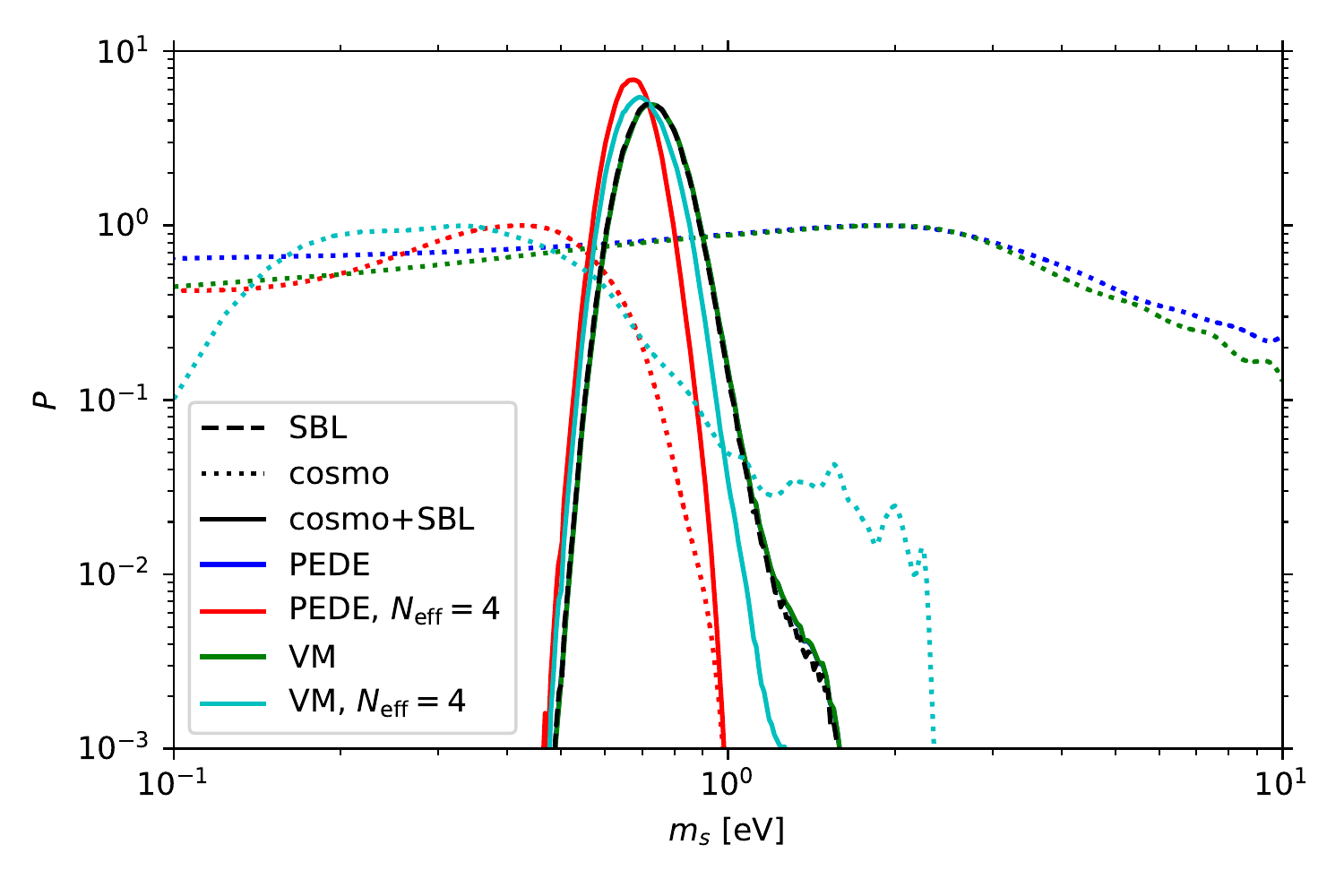}
    \caption{Posterior probability distribution of $m_s$ obtained from SBL appearance experiments (dashed), cosmological analyses (dotted) and their combination (solid)
    for a variety of cosmological models:
    PEDE with free \Neff\ (blue),
    PEDE with $\Neff=4$ (red),
    VM with free \Neff\ (green) and
    VM with $\Neff=4$ (cyan).}
    \label{fig:sbl_cosmo_combination}
\end{figure}

Still referring to the results obtained within the PEDE framework, Tab.~\ref{tab:PEDE2} reports the
constraints for the combined datasets, namely, CMB+Lensing+BAO+Pantheon and  CMB+Lensing+BAO+Pantheon+SBL for both a free value of $\Neff$ and for the $\Neff=4$ case. The latter case (i.e.\ $\Neff=4$) refers to a fourth, fully thermalized, sterile neutrino state. These bounds can be compared to those resulting within the canonical $\Lambda$CDM picture, shown in Tab.~\ref{tab:LCDM}.
Notice that the results we obtain for the case of a freely varying $\Neff$ are almost similar to those discussed above,
being the sterile neutrino mass unconstrained at 95\% CL for the dataset CMB+Lensing+BAO+Pantheon.
When combining this set of cosmological measurements with appearance SBL data, the limit on the sterile neutrino mass $m_s$ becomes tighter ($m_{s} = 0.75^{+0.06}_{-0.10}$ eV at 68\% CL for CMB+Lensing+BAO+Pantheon+SBL) and coincides with the region favoured by appearance SBL neutrino oscillation experiments,
i.e.\ close to $\Delta m^2_{41}\sim 0.6$~eV$^2$~\cite{Dentler:2018sju,steftalk}; see the blue solid curve in Fig.~\ref{fig:sbl_cosmo_combination}.
In such a case, the tension between SBL and cosmological estimates of the sterile neutrino mass is significantly alleviated.

Furthermore, we also notice that the value of the Hubble constant is considerably larger in the PEDE scheme than in the $\Lambda$CDM case: for CMB+Lensing+BAO+Pantheon dataset, we have $H_0= 70.32 _{-0.69}^{+0.78}$~km s$^{−1}$ Mpc$^{−1}$ for PEDE versus $H_0= 67.60 _{-0.51}^{+0.46}$~km s$^{−1}$ Mpc$^{−1}$ for $\Lambda$CDM, both at $68\%$~CL, while for the CMB+Lensing+BAO+Pantheon+SBL dataset, we obtain $H_0= 70.52 _{-0.80}^{+0.89}$~km s$^{−1}$ Mpc$^{−1}$ for PEDE versus $H_0= 67.69 _{-0.45}^{+0.45}$~km s$^{−1}$ Mpc$^{−1}$ for $\Lambda$CDM, both at $68\%$~CL.
The large value of the Hubble constant obtained within the PEDE \emph{minimal} dark energy scenario is depicted in Fig.~\ref{fig:3D}. The upper panel shows in red and blue lines the $68\%$ and $95\%$~CL contours in the $m_s-H_0$ plane arising from the analyses of CMB+Lensing+BAO+Pantheon data in the $\Lambda$CDM and PEDE frameworks, respectively.

When the fourth sterile neutrino is assumed to be fully thermal, we notice significant changes in the constraints with respect to the case of a free \Neff.
For instance, one can see that for CMB+Lensing+BAO+Pantheon, we have a $95\%$~CL bound on $m_s$ in the PEDE scenario ($m_s<0.65$~eV while it is unconstrained with free \Neff) which is still less tight than within the $\Lambda$CDM case ($m_s<0.255$~eV).
For CMB+Lensing+BAO+Pantheon+SBL, we get a tight constraint on $m_s$ leading to $m_s = 0.678^{+0.066}_{-0.052}$ eV (at 68\% CL), slightly higher than the $\Lambda$CDM prediction $m_s = 0.621^{+0.048}_{-0.056}$ eV (68\% CL).
Such results are shown in Fig.~\ref{fig:sbl_cosmo_combination}.  
When SBL data are added into the analyses, the limit barely changes.
Interestingly, the cosmological limit of $m_s<0.65$~eV lies very close to the best fit from SBL appearance data, which points to $\Delta m^2_{41}\simeq 0.55$~eV$^2$~\cite{Dentler:2018sju,steftalk}.
As in the case of a freely varying $N_{\rm{eff}}$, the addition of SBL measurements fixes a lower bound for $m_s$ of $\sim 0.5$~eV.
In this $N_{\rm{eff}}=4$ scenario, the $H_0$ tension is strongly lifted, as the value obtained for the Hubble constant within the PEDE cosmology is $H_0= 74.29 _{-0.85}^{+0.79}$~km s$^{−1}$ Mpc$^{−1}$ at 68$\%$~CL (for CMB+Lensing+BAO+Pantheon) and $H_0= 73.30 _{-0.57}^{+0.56}$~km s$^{−1}$ Mpc$^{−1}$ at 68$\%$~CL (for CMB+Lensing+BAO+Pantheon+SBL).
The effect of both a larger value of $m_s$ and $H_0$ obtained within the PEDE cosmology with respect to the canonical $\Lambda$CDM picture in the fully thermalized case is depicted in the lower panel of Fig.~\ref{fig:3D}, where one can clearly notice that the green contours, corresponding to the PEDE \emph{minimal} dark energy scenario, are shifted towards larger values of both $m_s$ and $H_0$.

\begingroup
\squeezetable
\begin{center}
\begin{table*}
\begin{tabular}{ccccccccccc}
\hline\hline
Parameters & CMB & CMB+BAO & CMB+Lensing & CMB+Pantheon & CMB+R20  \\ \hline

$\Omega_b h^2$ & $    0.02242_{-    0.00019-    0.00035}^{+    0.00017+    0.00039}$ & $    0.02234_{-    0.00020-    0.00034}^{+    0.00015+    0.00037}$  & $ 0.02242_{-    0.00017-    0.00033}^{+    0.00017+    0.00034}$ & $    0.02229_{-    0.00026-    0.00049}^{+    0.00026+    0.00053}$ & $    0.02238_{-    0.00022-    0.00039}^{+    0.00018+    0.00041}$  \\

$100\theta_{MC}$ & $    1.04062_{-    0.00037-    0.00075}^{+    0.00037+    0.00071}$ & $    1.04050_{-    0.00029-    0.00055}^{+    0.00029+    0.00057}$ & $    1.04072_{-    0.00034-    0.00068}^{+    0.00034+    0.00066}$ & $    1.03973_{-    0.00034-    0.00067}^{+    0.00034+    0.00066}$  & $    1.04038_{-    0.00032-    0.00063}^{+    0.00032+    0.00066}$  \\

$\tau$ & $    0.0527_{-    0.0078-    0.015}^{+    0.0078+    0.016}$ & $    0.0521_{-    0.0078-    0.015}^{+    0.0078+    0.016}$  & $    0.0518_{-    0.0079-    0.016}^{+    0.0079+    0.016}$  & $    0.0494_{-    0.0078-    0.015}^{+    0.0078+    0.016}$ &  $    0.0516_{-    0.0077-    0.015}^{+    0.0077+    0.016}$  \\

$n_s$ & $    0.9658_{-    0.0068-    0.012}^{+    0.0045+    0.013}$ & $    0.9633_{-    0.0068-    0.011}^{+    0.0044+    0.012}$ & $    0.9658_{-    0.0056-    0.010}^{+    0.0044+    0.011}$   & $    0.9634_{-    0.0096-    0.018}^{+    0.0096+    0.018}$ & $    0.9648_{-    0.0082-    0.012}^{+    0.0051+    0.014}$ \\

${\rm{ln}}(10^{10} A_s)$ & $    3.046_{-    0.016-    0.031}^{+    0.016+    0.033}$ & $    3.047_{-    0.016-    0.032}^{+    0.016+    0.033}$  & $    3.041_{-    0.016-    0.031}^{+    0.016+    0.031}$  & $    3.052_{-    0.017-    0.033}^{+    0.017+    0.035}$  & $    3.048_{-    0.016-    0.032}^{+    0.016+    0.034}$  \\

$M_{\rm vacuum}$ & $    0.925_{-    0.007-    0.025}^{+    0.014+    0.020}$ & $    0.9169_{-    0.0028-    0.0063}^{+    0.0034+    0.0060}$ & $    0.930_{-    0.005-    0.022}^{+    0.012+    0.016}$   & $    0.841_{-    0.013-    0.025}^{+    0.013+    0.025}$ & $    0.9112_{-    0.0048-    0.011}^{+    0.0058+    0.010}$ \\

$\Omega_{m0}$ & $    0.243_{-    0.029-    0.042}^{+    0.015+    0.051}$ & $0.2608_{-    0.0090-    0.015}^{+    0.0069+    0.016}$ & $    0.234_{-    0.026-    0.035}^{+    0.012+    0.047}$ & $    0.401_{-    0.021-    0.043}^{+    0.021+    0.042}$  & $    0.271_{-    0.012-    0.020}^{+    0.010+    0.022}$  \\

$\sigma_8$ & $    0.888_{-    0.021-    0.079}^{+    0.048+    0.062}$ & $    0.861_{-    0.033-    0.078}^{+    0.044+    0.071}$ & $    0.884_{-    0.022-    0.071}^{+    0.043+    0.057}$  & $    0.720_{-    0.037-    0.069}^{+    0.037+    0.073}$  & $    0.858_{-    0.032-    0.072}^{+    0.041+    0.068}$  \\

$H_0$ [km/s/Mpc]& $   77.8_{-    2.9-    7.1}^{+    3.9+    6.5}$ & $   75.14_{-    0.93-    1.9}^{+    0.93+    1.7}$  & $   79.3_{-    2.4-    6.8}^{+    3.8+    5.6}$ & $   62.1_{-    1.6-    2.8}^{+    1.4+    3.1}$  & $   73.7_{-    1.2-    2.5}^{+    1.2+    2.4}$\\

$m_{light}$  [eV]& $  < 0.0594 < 0.141 $ & $ 0.085^{+0.034}_{-0.071} < 0.181 $ & $ < 0.0680 < 0.144 $  & $ 0.242_{- 0.085-0.16}^{+    0.085+0.17}$ & $ 0.080_{- 0.067}^{+    0.032},  < 0.170 $ \\

$\Neff$ & $ < 3.19 < 3.41 $ & $ < 3.21 < 3.36 $ & $ < 3.15 < 3.31 $  & $ 3.44^{-0.20}_{+0.20} < 3.78  $   & $ < 3.29 < 3.49 $  \\

$m_s$  [eV]& $ < 2.54, \,\mbox{unconstr.} $ & $ < 2.20 < 6.62 $ & $ < 3.26 < 8.21 $  & $ 1.09^{+0.29+1.1}_{-0.69-0.9} $   & $ <1.78 < 5.95 $  \\

$S_8$ & $    0.797_{-    0.029-    0.055}^{+    0.029+    0.057}$ & $    0.802_{-    0.024-    0.057}^{+    0.032+    0.053}$ & $    0.778_{-    0.018-    0.037}^{+    0.018+    0.034}$  & $    0.831_{-    0.034-    0.068}^{+    0.034+    0.066}$  & $    0.816_{-    0.026-    0.058}^{+    0.029+    0.052}$  \\

$r_{\rm{drag}}$ [Mpc]& $  145.7_{-    0.5-    2.5}^{+    1.4+    1.7}$ & $  145.5_{-    0.6-    1.7}^{+    1.1+    1.4}$ & $  146.2_{-    0.4-    1.9}^{+    1.0+    1.3}$ & $  142.3_{-    1.7-    3.2}^{+    1.7+    3.1}$ & $ 144.8_{-    0.8-    2.4}^{+    1.7+    2.0}$ \\

$r_{\rm drag} h$ [Mpc]& $  113.4_{-    4.7-    12}^{+    6.4+    11}$ & $  109.3_{-    1.2-    2.3}^{+    1.2+    2.3}$  & $  115.9_{-    3.8-    11}^{+    6.1+    8.9}$  & $  88.4_{-    2.5-    4.4}^{+    2.2+    5.0}$  & $  106.8_{-    2.2-    4.3}^{+    2.2+    4.2}$  \\

\hline\hline
\end{tabular}
\caption{
Constraints on various free and derived parameters within the VM scenario. We report 68\% and 95\% CL limits. When the upper limit coincides with the upper prior range (see Tab.~\ref{tab:priors}), as in the case of $m_s$, we mark the parameter as ``unconstrained''.
}\label{tab:VM1}

\end{table*}
\end{center}
\endgroup


\begingroup
\squeezetable
\begin{center}
\begin{table*}
\begin{tabular}{ccccccccccc}
\hline\hline
Parameters & CMB+Lensing+BAO&~~~ CMB+Lensing+BAO & CMB+Lensing+BAO&~~~ CMB+Lensing+BAO \\
& +Pantheon ($\Neff$ free) & +Pantheon+SBL ($\Neff$ free) &~~~ +Pantheon ($\Neff = 4$) &~~~ +Pantheon+SBL ($\Neff = 4$)  \\
\hline
$\Omega_b h^2$ & $    0.02219_{-    0.00016-    0.00028}^{+    0.00013+    0.00029}$ & $    0.02223_{-    0.00015-    0.00029}^{+    0.00015+    0.00032}$ & $    0.02338_{-    0.00013-    0.00025}^{+    0.00013+    0.00025}$ & $    0.02338_{-    0.00013-    0.00024}^{+    0.00013+    0.00025}$ \\

$100\theta_{MC}$ & $    1.04043_{-    0.00029-    0.00057}^{+    0.00029+    0.00056}$ & $    1.04044_{-    0.00029-    0.00058}^{+    0.00029+    0.00058}$ & $    1.03993_{-    0.00028-    0.00056}^{+    0.00028+    0.00055}$ & $    1.03996_{-    0.00028-    0.00055}^{+    0.00028+    0.00055}$ \\

$\tau$ & $    0.0528_{-    0.0076-    0.015}^{+    0.0076+    0.016}$ & $    0.0532_{-    0.0079-    0.015}^{+    0.0069+    0.017}$ & $    0.0604_{-    0.0095-    0.017}^{+    0.0077+    0.018}$ & $    0.0636_{-    0.0098-    0.017}^{+    0.0080+    0.019}$\\

$n_s$ & $    0.9593_{-    0.0043-    0.0084}^{+    0.0043+    0.0084}$ & $    0.9625_{-    0.0051-    0.0086}^{+    0.0039+    0.0093}$ & $    1.0061_{-    0.0025-    0.015}^{+    0.0050+    0.010}$ & $    1.0054_{-    0.0037-    0.0073}^{+    0.0037+    0.0074}$ \\

${\rm{ln}}(10^{10} A_s)$ & $    3.046_{-    0.015-    0.031}^{+    0.015+    0.032}$ & $    3.046_{-    0.016-    0.030}^{+    0.016+    0.033}$ & $    3.073_{-    0.019-    0.034}^{+    0.015+    0.036}$ & $    3.079_{-    0.019-    0.033}^{+    0.016+    0.037}$  \\

$M_{\rm vacuum}$ & $    0.9090_{-    0.0031-    0.0062}^{+    0.0031+    0.0060}$  & $    0.9097_{-    0.0031-    0.0064}^{+    0.0031+    0.0059}$ & $    0.9170_{-    0.0024-    0.0077}^{+    0.0039+    0.0067}$ & $    0.9155_{-    0.0032-    0.0055}^{+    0.0026+    0.0062}$ \\

$\Omega_{m0}$ & $    0.2801_{-    0.0071-    0.014}^{+    0.0071+    0.014}$ & $    0.2774_{-    0.0072-    0.014}^{+    0.0072+    0.014}$ & $    0.2515_{-    0.0097-    0.016}^{+    0.0049+    0.021}$ & $    0.2557_{-    0.0049-    0.016}^{+    0.0079+    0.012}$ \\

$\sigma_8$ & $    0.809_{-    0.021-    0.047}^{+    0.025+    0.044}$ & $    0.822_{-    0.020-    0.039}^{+    0.020+    0.039}$ &  $    0.846_{-    0.016-    0.14}^{+    0.045+    0.07}$ &  $    0.823_{-    0.031-    0.042}^{+    0.008+    0.079}$ \\

$H_0$ [km/s/Mpc]& $   73.03_{-    0.80-    1.6}^{+    0.80+    1.6}$ & $   73.10_{-    0.82-    1.6}^{+    0.82+    1.6}$ & $   77.18_{-    0.80-    1.6}^{+    0.80+    1.5}$ & $   76.86_{-    0.81-    1.4}^{+    0.70+    1.5}$  \\

$m_{light}$ [eV]& $ 0.166_{-    0.039 -0.080}^{+    0.039+0.070} $ & $ 0.162_{-    0.035 -0.068}^{+    0.035+0.066} $ & $   < 0.0417 < 0.0854 $ & $ <0.0259 <0.0771 $ \\

$\Neff$ & $ < 3.12 < 3.24 $ & $ < 3.14 < 3.25 $ & [$4$] & [$4$] \\

$m_s$ [eV]& $ <4.2 < 8.53    $ &$0.79^{+0.07+0.020}_{-0.011-0.018}$ & $ 0.49^{+0.05}_{-0.33} < 1.30 $ &$0.64^{+0.19+0.28}_{-0.02-0.53}$ \\

$S_8$ & $    0.782_{-    0.014-    0.035}^{+    0.019+    0.032}$ & $    0.791_{-    0.014-    0.028}^{+    0.014+    0.027}$ & $    0.774_{-    0.011-    0.086}^{+    0.030+    0.047}$ & $    0.759_{-    0.024-    0.031}^{+    0.009+    0.052}$ \\

$r_{\rm{drag}}$ [Mpc]& $  146.01_{-    0.22-    1.4}^{+    0.72+    0.9}$ & $  145.99_{-    0.31-    1.2}^{+    0.70+    0.9}$ & $  139.16_{-    0.13-    0.76}^{+    0.24+    0.48}$ & $  139.12_{-    0.17-    0.32}^{+    0.17+    0.33}$ \\

$r_{\rm drag} h$ [Mpc]& $  106.6_{-    1.1-    2.1}^{+    1.1+    2.2}$ & $  106.7_{-    1.1-    2.2}^{+    1.1+    2.2}$ & $  107.4_{-    1.1-    2.6}^{+    1.3+    2.3}$ & $  106.9_{-    1.2-    2.0}^{+    1.0+    2.2}$  \\
\hline\hline
\end{tabular}
\caption{Constraints on various free and derived parameters within the VM scenario, the combined datasets CMB+Lensing+BAO+Pantheon and CMB+Lensing+BAO+Pantheon+SBL. We present the cases where $\Neff$ is either freely varying (second and third columns), or fixed to $\Neff = 4$ (fourth and fifth columns).}\label{tab:VM2}
\end{table*}
\end{center}
\endgroup

Table~\ref{tab:VM1} shows the constraints for the Vacuum Metamorphosis (VM) model described in the previous section for several data combinations.
Notice that, as in the case of the PEDE scenario, the bounds on both the sterile neutrino and the active neutrino masses are significantly relaxed.
Again, this is due to the large negative value adopted by the effective dark energy equation of state within the VM framework, $w_{\rm VMDE}$, especially at low redshifts; see Eq.~(\ref{eq:wpde}) and Refs.~\cite{DiValentino:2017rcr,DiValentino:2020kha,DiValentino:2021zxy}.
These findings agree with previous results in the literature; see e.g.\  Ref.~\cite{DiValentino:2021zxy}, in which the authors found a $3\sigma$ evidence for a non-zero neutrino mass for the Planck+BAO+Pantheon dataset combination, always in the context of the VM scenario.
Similarly also to the PEDE model, the values of $N_{\rm{eff}}$ are barely changed with respect to the $\Lambda$CDM scheme, so that also in this case a suppression mechanism to avoid a complete sterile neutrino thermalization in the early universe would be required.
As an example, $N_{\rm{eff}}<3.36$~eV at $95\%$~CL for the CMB plus BAO data analyses.
The tension on the Hubble constant is notably reduced for almost all the data combinations shown in Tab.~\ref{tab:VM1}: for Planck 2018 CMB+Lensing, $H_0= 79.3_{-2.4}^{+3.8}$~km s$^{−1}$ Mpc$^{−1}$ at 68$\%$~CL.
This resolution of the Hubble constant tension was also present in the PEDE model.
However, the error bars now are much larger, due to the strong degeneracy between $H_0$ and the parameter $M$, which also depends on the $H_0$ value, as we discussed in Section~\ref{ssec:VM}.

Instead, for the CMB+Pantheon data combination, we find a very low value of $H_0$ and a relatively low, non-zero preferred value for the sterile neutrino mass ($m_s=1.09_{-0.69}^{+0.29}$~eV at $68\%$~CL).
These results agree with those of Ref.~\cite{DiValentino:2021zxy} and are due to the fact that the addition of SNIa no longer allows $w_{\rm VMDE}$ to adopt very larger negative values at low redshifts.
The overall effect is translated into both a smaller Hubble constant and a smaller sterile neutrino mass.

Table~\ref{tab:VM2} shows the results within the VM \emph{minimal} cosmology for the combined datasets, i.e.\ CMB+Lensing+BAO+Pantheon and CMB+Lensing+BAO+Pantheon+SBL, considering the two possible scenarios, i.e.\ when \Neff\ is treated as a free parameter and when $\Neff = 4$, referring to the case with a fourth, fully thermalized, sterile
neutrino state.
For both the choices of \Neff, the results on $m_s$ for cosmological data only and for the combined cosmological plus SBL data analyses are depicted in Fig.~\ref{fig:sbl_cosmo_combination}; see the dotted and solid green lines, respectively.
The results for the case of a freely varying $\Neff$ are almost identical to the PEDE case,
except for the active neutrino mass and the Hubble constant, whose bounds and mean values are less constraining and higher than within the PEDE scheme, respectively.
Notice again that the sterile neutrino mass is unconstrained at 95\%~CL when $\Neff$ is free and considering the case CMB+Lensing+BAO+Pantheon.
For CMB+Lensing+BAO+Pantheon+SBL, $m_s$ is constrained to $m_s = 0.79^{+0.07}_{-0.011}$~eV (at 68\% CL).
The case of $\Neff=4$, instead, has the $95\%$~CL upper cosmological bound on $m_s$ equal to $1.3$~eV.
This value is in perfect agreement with the bounds obtained from SBL appearance experiments, see the dotted cyan lines in Fig.~\ref{fig:sbl_cosmo_combination} and compare them with the dashed black line.
Notice that within the VM dark energy framework with $\Neff=4$, the combination with SBL data \emph{tightens} the cosmological upper limit; see the solid cyan line in Fig.~\ref{fig:sbl_cosmo_combination}.
The value of the Hubble constant for the VM scenario with $\Neff=4$ is $H_0=77.18 \pm 0.80$~km s$^{−1}$ Mpc$^{−1}$ at 68$\%$~CL.

\begingroup
\squeezetable
\begin{center}
\begin{table*}
\begin{tabular}{ccccccccccc}
\hline\hline
Parameters & CMB & CMB+BAO & CMB+Lensing & CMB+Pantheon & CMB+R20  \\ \hline

$\Omega_c h^2$ & $    0.1201_{-    0.0016-    0.0038}^{+    0.0016+    0.0040}$ & $    0.1207_{-    0.0017-    0.0043}^{+    0.0020+    0.0037}$ & $    0.1199_{-    0.0018-    0.0035}^{+    0.0018+    0.0037}$ & $    0.1217_{-    0.0024-    0.0046}^{+    0.0021+    0.0052}$ & $    0.1205_{-    0.0020-    0.0041}^{+    0.0020+    0.0040}$ \\

$\Omega_b h^2$ & $    0.02239_{-    0.00016-    0.00030}^{+    0.00016+    0.00032}$ & $    0.02236_{-    0.00015-    0.00030}^{+    0.00015+    0.00031}$  & $ 0.02241_{-    0.00015-    0.00031}^{+    0.00015+    0.00030}$ & $    0.02235_{-    0.00018-    0.00034}^{+    0.00018+    0.00035}$ & $    0.02235_{-    0.00016-    0.00031}^{+    0.00016+    0.00033}$  \\

$100\theta_{MC}$ & $    1.04077_{-    0.00034-    0.00067}^{+    0.00034+    0.00065}$ & $    1.04070_{-    0.00031-    0.00060}^{+    0.00031+    0.00062}$ & $    1.04081_{-    0.00034-    0.00068}^{+    0.00034+    0.00065}$ & $    1.04058_{-    0.00037-    0.00075}^{+    0.00037+    0.00070}$  & $    1.04071_{-    0.00034-    0.00067}^{+    0.00034+    0.00067}$  \\

$\tau$ & $    0.0545_{-    0.0078-    0.015}^{+    0.0078+    0.016}$ & $    0.0539_{-    0.0079-    0.015}^{+    0.0079+    0.016}$  & $    0.0541_{-    0.0075-    0.015}^{+    0.0075+    0.015}$  & $    0.0542_{-    0.0079-    0.015}^{+    0.0079+    0.016}$ &  $    0.0544_{-    0.0078-    0.015}^{+    0.0078+    0.016}$  \\

$n_s$ & $    0.9647_{-    0.0049-    0.009}^{+    0.0049+    0.010}$ & $    0.9630_{-    0.0049-    0.0087}^{+    0.0042+    0.0098}$ & $    0.9649_{-    0.0047-    0.0092}^{+    0.0047+    0.0093}$   & $    0.9627_{-    0.0058-    0.011}^{+    0.0050+    0.011}$ & $    0.9631_{-    0.0048-    0.0091}^{+    0.0048+    0.0097}$ \\

${\rm{ln}}(10^{10} A_s)$ & $    3.046_{-    0.016-    0.031}^{+    0.016+    0.034}$ & $    3.048_{-    0.016-    0.032}^{+    0.016+    0.032}$  & $    3.045_{-    0.015-    0.029}^{+    0.015+    0.031}$  & $    3.050_{-    0.018-    0.032}^{+    0.016+    0.036}$  & $    3.048_{-    0.016-    0.031}^{+    0.016+    0.033}$  \\

$M_{\rm vacuum}$ & $    0.910_{-    0.011-    0.019}^{+    0.010+    0.024}$ & $    0.8948_{-    0.0047-    0.0064}^{+    0.0021+    0.0091}$ & $    0.917_{-    0.008-    0.023}^{+    0.017+    0.019}$   & $    0.8903_{-    0.0032-    0.0066}^{+    0.0032+    0.0064}$ & $    0.9001_{-    0.0079-    0.011}^{+    0.0049+    0.013}$ \\

$\Omega_{m0}$ & $    0.254_{-    0.012-    0.036}^{+    0.016+    0.028}$ & $0.2722_{-    0.0064-    0.012}^{+    0.0057+    0.012}$ & $    0.245_{-    0.020-    0.029}^{+    0.015+    0.032}$ & $    0.283_{-    0.015-    0.021}^{+    0.009+    0.026}$  & $    0.2672_{-    0.0099-    0.018}^{+    0.0081+    0.019}$  \\

$\sigma_8$ & $    0.858_{-    0.027-    0.066}^{+    0.034+    0.064}$ & $    0.834_{-    0.020-    0.053}^{+    0.030+    0.047}$ & $    0.867_{-    0.025-    0.059}^{+    0.034+    0.055}$  & $    0.812_{-    0.026-    0.077}^{+    0.047+    0.064}$  & $    0.839_{-    0.020-    0.061}^{+    0.034+    0.053}$  \\

$H_0$ [km/s/Mpc]& $   75.9_{-    2.2-    3.3}^{+    1.3+    5.0}$ & $   73.42_{-    0.67-    1.2}^{+    0.52+    1.3}$  & $   77.1_{-    2.4-    4.2}^{+    2.9+    4.2}$ & $   72.6_{-    0.7-    2.1}^{+    1.1+    1.5}$  & $   74.1_{-    1.0-    1.7}^{+    0.8+    2.1}$\\

$m_{light}$ [eV]& $  < 0.0410 < 0.105 $ & $ <0.0479 < 0.0964 $ & $ < 0.0435 < 0.0965 $  & $ <0.0738 <0.170$ & $ <0.0542  < 0.123 $ \\

$\Neff$ & $ < 3.11 < 3.24 $ & $ < 3.13 < 3.25 $ & $ < 3.11 < 3.21 $  & $ <3.19 < 3.39  $   & $ < 3.11 < 3.24 $  \\

$m_s$ [eV]& $ < 5.05, \, \mbox{unconstr.} $ & $ < 4.70, \,\mbox{unconstr.} $ & $ < 4.72, \, \mbox{unconstr.} $  & $ <3.52 <8.01 $   & $ <4.97, \,\mbox{unconstr.} $  \\

$S_8$ & $    0.788_{-    0.018-    0.044}^{+    0.023+    0.042}$ & $    0.795_{-    0.019-    0.047}^{+    0.026+    0.043}$ & $    0.783_{-    0.016-    0.039}^{+    0.020+    0.033}$  & $    0.787_{-    0.022-    0.055}^{+    0.031+    0.049}$  & $    0.792_{-    0.019-    0.048}^{+    0.026+    0.044}$  \\

$r_{\rm{drag}}$ [Mpc]& $  146.44_{-    0.31-    1.5}^{+    0.79+    1.2}$ & $  146.17_{-    0.34-    1.5}^{+    0.81+    1.1}$ & $  146.48_{-    0.33-    1.4}^{+    0.74+    1.1}$ & $  145.6_{-    0.5-    2.1}^{+    1.4+    1.5}$ & $ 146.30_{-    0.31-    1.6}^{+    0.82+    1.2}$ \\

$r_{\rm drag} h$ [Mpc]& $  111.1_{-    3.4-    5.3}^{+    2.0+    7.7}$ & $  107.32_{-    0.99-    1.8}^{+    0.87+    2.0}$  & $  112.9_{-    3.6-    6.4}^{+    3.6+   6.3}$  & $  105.7_{-    1.3-    3.3}^{+    1.9+    2.9}$  & $  108.4_{-    1.6-    3.0}^{+    1.6+    3.2}$  \\

\hline\hline
\end{tabular}
\caption{
Constraints on various free and derived parameters within the EVM scenario. We report 68\% and 95\% CL limits. When the upper limit coincides with the upper prior range (see Tab.~\ref{tab:priors}), as in the case of $m_s$, we mark the parameter as ``unconstrained''.
}\label{tab:VMext}

\end{table*}
\end{center}
\endgroup


\begingroup
\squeezetable
\begin{center}
\begin{table*}
\begin{tabular}{ccccccccccc}
\hline\hline                                 Parameters & CMB+Lensing+BAO&~~~ CMB+Lensing+BAO & CMB+Lensing+BAO&~~~ CMB+Lensing+BAO \\
& +Pantheon ($\Neff$ free) & +Pantheon+SBL ($\Neff$ free) &~~~ +Pantheon ($\Neff = 4$) &~~~ +Pantheon+SBL ($\Neff = 4$)  \\
\hline

$\Omega_c h^2$ & $    0.1214_{-    0.0021-    0.0044}^{+    0.0019+    0.0047}$ & $    0.1233_{-    0.0020-    0.0033}^{+    0.0015+    0.0036}$ & $    0.1348_{-    0.0012-    0.0022}^{+    0.0012+    0.0022}$ & $    0.1337_{-    0.0012-    0.0024}^{+    0.0012+    0.0024}$ \\

$\Omega_b h^2$ & $    0.02238_{-    0.00016-    0.00030}^{+    0.00016+    0.00031}$ & $    0.02241_{-    0.00016-    0.00030}^{+    0.00016+    0.00032}$ & $    0.02301_{-    0.00014-    0.00027}^{+    0.00014+    0.00028}$ & $    0.02299_{-    0.00014-    0.00028}^{+    0.00014+    0.00028}$ \\

$100\theta_{MC}$ & $    1.04065_{-    0.00030-    0.00066}^{+    0.00033+    0.00061}$ & $    1.04051_{-    0.00034-    0.00068}^{+    0.00034+    0.00065}$ & $    1.03944_{-    0.00029-    0.00056}^{+    0.00029+    0.00056}$ & $    1.03949_{-    0.00029-    0.00056}^{+    0.00029+    0.00057}$ \\

$\tau$ & $    0.0544_{-    0.0075-    0.014}^{+    0.0075+    0.016}$ & $    0.0553_{-    0.0076-    0.015}^{+    0.0076+    0.015}$ & $    0.0629_{-    0.0093-    0.015}^{+    0.0077+    0.018}$ & $    0.0687_{-    0.0099-    0.017}^{+    0.0084+    0.019}$\\

$n_s$ & $    0.9637_{-    0.0057-    0.010}^{+    0.0043+    0.011}$ & $    0.9675_{-    0.0055-    0.010}^{+    0.0047+    0.010}$ & $    0.9950_{-    0.0038-    0.0074}^{+    0.0038+    0.0072}$ & $    0.9922_{-    0.0039-    0.0078}^{+    0.0039+    0.0075}$ \\

${\rm{ln}}(10^{10} A_s)$ & $    3.050_{-    0.015-    0.029}^{+    0.015+    0.031}$ & $    3.054_{-    0.016-    0.030}^{+    0.016+    0.032}$ & $    3.094_{-    0.018-    0.030}^{+    0.015+    0.034}$ & $    3.103_{-    0.019-    0.032}^{+    0.016+    0.036}$  \\

$M_{\rm vacuum}$ & $    0.8909_{-    0.0021-    0.0042}^{+    0.0021+    0.0045}$  & $    0.8902_{-    0.0022-    0.0045}^{+    0.0022+    0.0045}$ & $    0.8809_{-    0.0023-    0.0035}^{+    0.0012+    0.0044}$ & $    0.8800_{-    0.0034-    0.0045}^{+    0.0013+    0.0064}$ \\

$\Omega_{m0}$ & $    0.2770_{-    0.0063-    0.011}^{+    0.0055+    0.012}$ & $    0.2769_{-    0.0060-    0.010}^{+    0.0050+    0.012}$ & $    0.2778_{-    0.0064-    0.011}^{+    0.0056+    0.012}$ & $    0.2913_{-    0.0050-    0.010}^{+    0.0050+    0.010}$ \\

$\sigma_8$ & $    0.828_{-    0.018-    0.045}^{+    0.026+    0.040}$ & $    0.840_{-    0.014-    0.029}^{+    0.016+    0.028}$ &  $    0.844_{-    0.015-    0.037}^{+    0.021+    0.033}$ &  $    0.789_{-    0.012-    0.024}^{+    0.012+    0.022}$ \\

$H_0$ [km/s/Mpc]& $   73.03_{-    0.53-    1.1}^{+    0.53+    1.1}$ & $   73.26_{-    0.46-    1.1}^{+    0.57+    1.0}$ & $   76.27_{-    0.51-    1.1}^{+    0.56+    1.0}$ & $   75.12_{-    0.47-    0.84}^{+    0.38+    0.91}$  \\

$m_{light}$ [eV]& $ <0.0475<0.0877 $ & $ <0.0457<0.0863 $ & $   < 0.0313 < 0.0647 $ & $   < 0.0157 < 0.0341 $ \\

$\Neff$ & $ < 3.18 < 3.34 $ & $ < 3.26 < 3.42 $ & [$4$] & [$4$] \\

$m_s$ [eV]& $ <3.62 < 8.50    $ &$0.749^{+0.064+0.17}_{-0.096-0.16}$ & $ 0.25^{+0.04}_{-0.15} < 0.467 $ &$0.669^{+0.051+0.13}_{-0.071-0.12}$ \\

$S_8$ & $    0.795_{-    0.015-    0.039}^{+    0.022+    0.035}$ & $    0.807_{-    0.012-    0.024}^{+    0.012+    0.024}$ & $    0.812_{-    0.013-    0.030}^{+    0.015+    0.026}$ & $    0.778_{-    0.012-    0.024}^{+    0.012+    0.023}$ \\

$r_{\rm{drag}}$ [Mpc]& $  145.8_{-    0.5-    1.9}^{+    1.1+    1.4}$ & $  145.2_{-    0.8-    2.2}^{+    1.4+    1.8}$ & $  138.33_{-    0.22-    0.43}^{+    0.22+    0.43}$ & $  138.20_{-    0.23-    0.45}^{+    0.23+    0.44}$ \\

$r_{\rm drag} h$ [Mpc]& $  106.44_{-    0.83-    1.6}^{+    0.83+    1.6}$ & $  106.39_{-    0.74-    1.6}^{+    0.83+    1.5}$ & $  105.50_{-    0.82-    1.6}^{+    0.82+   1.5}$ & $  103.81_{-    0.72-    1.3}^{+    0.63+   1.4}$  \\
\hline\hline
\end{tabular}
\caption{Constraints on various free and derived parameters within the EVM scenario, considering the combined datasets CMB+Lensing+BAO+Pantheon and CMB+Lensing+BAO+Pantheon+SBL. We present the cases where $\Neff$ is either freely varying (second and third columns), or fixed to $\Neff = 4$ (fourth and fifth columns). }\label{tab:VMext2}
\end{table*}
\end{center}
\endgroup

Even if it is not among the main focus of this study, as it is not a \emph{minimal} dark energy scenario, however, we shall briefly comment on the constraints within the extended VM model, i.e.\ within the \emph{Elaborated Vacuum Metamorphosis} (EVM), where a cosmological constant at high redshifts is included as the vacuum expectation value of the massive scalar field.
Table~\ref{tab:VMext} shows the equivalent to Tab.~\ref{tab:VM1} but within this EVM model.
While the results concerning the sterile neutrino parameters and the Hubble constant are very similar to those previously shown in the original VM case, the CMB+Pantheon data analyses are very different, as in this case they offer the same relief to both the SBL-cosmological sterile neutrino and $H_0$ tensions, which is also visible in the remaining data combinations explored here.
This is due to the fact that the dark energy equation of state after the phase transition is now effectively a free parameter and therefore the addition of the Pantheon sample of the SNIa data no longer fixes this parameter.
Finally, Tab.~\ref{tab:VMext2} shows the constraints for the full data combination for both a freely varying value of $N_{\rm{eff}}$ and for $N_{\rm{eff}}=4$.
For the latter case, $m_s<0.467$~eV and $m_{light}< 0.0647$ both at $95\%$~CL and $H_0 = 76.27_{-0.51}^{+0.56}$~km s$^{−1}$ Mpc$^{−1}$ at 68$\%$~CL.

We close this section by reporting the results of a Bayesian model comparison analysis on the minimal dark energy scenarios considered here versus the $\Lambda$CDM case, considering both the freely varying $\Neff$ and the $\Neff=4$ cases,
for the complete dataset, namely, CMB+Lensing+BAO+Pantheon+SBL.
This is shown in Table~\ref{tab:BE} (see the last column), where we present the Bayes factors $\ln B_{ij}$ to estimate the strength of the preference for any cosmological model $M_i$ (PEDE, VM, EVM) with respect to the base $\Lambda$CDM model.
Notice that positive values favor the considered model, while negative ones prefer the $\Lambda$CDM case.
For the scenarios with $\Neff=4$, only PEDE offers an improvement with respect to the $\Lambda$CDM case when SBL appearance data are included together with the CMB+Lensing+BAO+Pantheon cosmological data.
This is a consequence of the very similar results on $m_s$ between the PEDE fit without SBL data and the SBL constraints alone.
In all the other cases, the $\Lambda$CDM model is favored with respect to the non-standard case.
Notice that, according to Table~\ref{tab:BE}, both the VM and EVM models are decisively disfavored by the considered data combinations, for both with free or fixed \Neff.
Therefore, the PEDE phenomenological dark energy model may provide a better description of a number of cosmological, astrophysical, and, eventually, particle physics observations than the standard $\Lambda$CDM model.

\begin{table*}
\begin{center}
\renewcommand{\arraystretch}{1}
\begin{tabular}{c|cccccccccc}
\hline
Models& CMB+Lensing+BAO& CMB+Lensing+BAO \\
 &  +Pantheon+SBL ($\Neff$ free) & +Pantheon+SBL ($\Neff = 4$)  \\

\hline \hline

        PEDE    &  $-3.7$    &  $+5.2$  \\

        VM      &  $-99.2$  &  $-92.0$  \\

        EVM   &  $-38.1$   &  $-19.3$  \\

        \hline
    \end{tabular}
    \caption{Natural logarithm of the Bayes factors between the PEDE, VM, EVM \emph{minimal} dark energy cosmologies and the $\Lambda$CDM case, considering the CMB+Lensing+BAO+Pantheon+SBL and either a free $\Neff$ or fixed $\Neff = 4$.
    Negative values prefer the $\Lambda$CDM scenario.}
    \label{tab:BE}
    \end{center}
\end{table*}

\section{Concluding remarks}
\label{sec-conclusion}

Tensions in the cosmological and neutrino oscillation measurements  may have a common explanation.
On the one hand, there is the well-known Hubble constant tension, a $4-6\sigma$ discrepancy between high and low redshift measurements of the Hubble constant.
On the other hand, Short Baseline (SBL) neutrino oscillation experiments require the existence of a fourth sterile neutrino with a $\sim$ eV mass. 
However, this preferred value for the sterile neutrino mass is at odds with cosmological bounds, unless some unknown mechanism suppresses the thermalization of the new neutrino in the early universe and consequently reduce its contribution to \Neff.
A sterile neutrino mass at the eV scale is indeed allowed when the thermalization leads to values of \Neff\ very close to the standard value with three neutrinos, 3.044.

The assumption of an underlying cosmological scenario may bias the results on the different parameters, even if the scenario is the canonical and minimal $\Lambda$CDM picture, which nevertheless provides an excellent fit to almost all available current cosmological observations.
It is therefore timely to investigate if there are alternative cosmological scenarios, as minimal as the $\Lambda$CDM one, which not only provide a very good fit to observations, but also alleviate current tensions.
In this regard, we have explored here two possible \emph{minimal} dark energy models, that do not involve additional parameters with respect to the simple $\Lambda$CDM scenario.
One of them is constructed from a phenomenological ground while the other one emerges from first principles.
We focused on the Phenomenologically Emergent Dark Energy (PEDE) and the Vacuum Metamorphosis (VM) models, which both modify only late time physics, and we have shown that such models can alleviate the current existing tensions in a significant way.
After dedicated analyses and fits to cosmological and SBL appearance measurements, we have found that in the two scenarios when we assume $\Neff \simeq 4$, the sterile neutrino mass bound is significantly relaxed, due to the large negative value that the effective dark energy equation of state is allowed to adopt in these \emph{minimal} cosmologies.
We have found that $m_s<0.65\ (1.3)$~eV at 95\% CL for the PEDE (VM) \emph{minimal} dark energy scenario with $\Neff=4$, respectively, when considering the most complete dataset combination.
Notice, moreover, that the PEDE \emph{minimal} scenario with $\Neff =4$ is the only case that is favored with respect to the $\Lambda$CDM model when the CMB+Lensing+BAO+Pantheon+SBL dataset is considered,
while the VM and EVM models are strongly disfavored by data.
When \Neff\ is free to vary, the sterile neutrino mass is basically unconstrained.
In such case, however, the low value of \Neff\ close to 3 cannot be explained by the preferred active-sterile neutrino oscillation parameters,
and some new mechanism or extra interaction is required to avoid the complete thermalization of the sterile neutrino state in the early universe.
Last, but not least, the Hubble constant tension is strongly relieved also in both \emph{minimal} dark energy scenarios.

With the growing sensitivity of the observational surveys that we witnessed over the past years, we are quite hopeful that future terrestrial searches for sterile neutrinos via oscillation experiments, and cosmological probes, testing the dark energy/radiation components and improving the current uncertainties on the Hubble parameter, will shed light on these tensions and also on the most favoured description of our universe.

\section*{Acknowledgments}
We thank the referee for some important comments which improved the article significantly.
EDV is supported by a Royal Society Dorothy Hodgkin Research Fellowship.
SG acknowledges financial support from the European Union's Horizon 2020 research and innovation programme under the Marie Skłodowska-Curie grant agreement No 754496 (project FELLINI).
CG is supported by the research grant "The Dark Universe: A Synergic Multimessenger Approach" number 2017X7X85K under the program PRIN 2017 funded by the Ministero dell'Istruzione, Universit\`a e della Ricerca (MIUR). 
OM is supported by the Spanish grants PID2020-113644GB-I00, PROMETEO/2019/083 and by the European ITN project HIDDeN (H2020-MSCA-ITN-2019//860881-HIDDeN). SP acknowledges the financial supports from the Science and Engineering Research Board, Govt. of India, under Mathematical Research Impact-Centric Support Scheme (File No. MTR/2018/000940) and the Department of Science and Technology (DST), Govt. of India, under the Scheme
``Fund for Improvement of S\&T Infrastructure (FIST)'' [File No. SR/FST/MS-I/2019/41].  WY was supported by the National Natural Science Foundation of China under Grants No. 12175096 and No. 11705079, and Liaoning Revitalization Talents Program under Grant no. XLYC1907098.


\bibliography{biblio}

\end{document}